# Beyond the Parasitic Limit: A Nanoprobing Framework for De-embedding Intrinsic Ferroelectric Properties at the Deep Sub-Micrometer Scale

**Authors:** Niklas Kyoushi[1,2], Georg Schönweger[2,3], Dirk Meyners[1], Ole Gronenberg[1], Adrian Petraru[3], Fabian Lofink[1,2], Simon Fichter[1,2]

[1]Department of Material Science, Kiel University, Kaiserstr. 2, D-24143 Kiel, Germany

[2]Fraunhofer Institute for Silicon Technology, Fraunhoferstr. 1, D-25524 Itzehoe, Germany

[3]Department of Electrical and Information Engineering, Kiel University, Kaiserstr. 2, D-24143 Kiel, Germany

**Abstract:** The continued scaling of ferroelectric devices is critical for next-generation computing architectures, yet it is fundamentally challenged by a metrological bottleneck: at the deep sub-micrometer scale, intrinsic material properties are heavily masked by extrinsic parasitic impedances and geometric fringing fields. Here, we introduce a quantitative, in-situ nanoprobing framework capable of resolving the true electrical response of ferroelectric capacitors down to 165 nm in diameter without the need for lithographic bond pads. Using 20 nm thick $Al_{0.72}Sc_{0.28}N$ as a model system, we establish a non-linear 'Screened Power Law' model to decouple attofarad-level device capacitances from massive near-field probe interactions. Furthermore, we demonstrate that the apparent degradation of dielectric loss at the nanoscale is a geometric dilution artifact, which we overcome through a conductance scaling analysis. Finally, we apply this framework to large-signal characterization, utilizing leakage-compensation and noise filtering protocols to extract pristine intrinsic hysteresis (C-V and J-E) loops in the discrete few-grain limit. These findings provide a universal analytical toolkit required to overcome the measurement limits of deep-submicron ferroelectrics.

# 1 Introduction

The scaling of ferroelectric thin films into the deep sub-micrometer regime is a fundamental prerequisite for high-density non-volatile memory and neuromorphic computing architectures. Wurtzite aluminum scandium nitride (AlScN) has emerged as a highly promising candidate for these applications due to its large remanent polarization, CMOS compatibility, and stable scaling behavior [1, 2]. However, as device dimensions contract below the 1 µm threshold, validating these intrinsic properties at the discrete capacitor level becomes a severe metrological challenge. At this scale, the physical measurement boundary conditions undergo a radical transition, causing the intrinsic electrical signal to drop below the background parasitic floor of conventional test setups.

Current characterization techniques fail to provide a complete, quantitative assessment of nanoscale ferroelectrics. Piezoresponse force microscopy (PFM) offers excellent spatial resolution and is routinely used to verify domain dynamics; however, it is inherently susceptible to non-local electrostatic forces, charge injection, and topographic crosstalk [3–6]. These artifacts can perfectly mimic ferroelectric hysteresis, obscuring the extraction of quantitative dielectric metrics and true coercive fields. Conversely, standard electrical probing utilizing lithographic bond pads introduces massive parasitic RC networks that completely mask the device response. Even direct electrical nanoprobing confronts the 'attofarad barrier,' where the intrinsic device capacitance ($C_{\text{device}} < 1\ \text{fF}$) is heavily shunted by hundreds of femtofarads of stray background generated by the macroscopic probe shaft and unshielded environment [7, 8]. Without analytical de-embedding, these near-field probe interactions, localized edge effects, and geometric current constrictions are frequently misinterpreted as intrinsic finite-size scaling limits or material degradation.

To overcome this measurement barrier, we present a comprehensive in-situ electrical nanoprobing framework capable of resolving the true dielectric and ferroelectric properties of isolated AlScN capacitors down to 165 nm in diameter, entirely bypassing the need for lithographic bond pads. By bridging five orders of magnitude in device area, spanning from macroscopic capacitors (Ø~50 µm) down to discrete nanostructural clusters of fewer than 30 grains, we systematically deconstruct the dominating parasitic regimes.

In this work, we first establish a physical 'Screened Power Law' capacitance model to quantify and de-embed the dynamic stray fields inherent to near-field probe measurements. Second, we deploy area-perimeter conductance scaling to separate intrinsic material damping from geometric dilution, recovering the true bulk dielectric loss. Finally, we apply this framework to large-signal characterization, utilizing numerical leakage-compensation protocols to extract pristine displacement currents and construct intrinsic J-E hysteresis loops in the deep sub-micrometer limit. This approach provides the quantitative analytical toolkit required to decouple fundamental material physics from extrinsic measurement artifacts in nanoscale ferroelectrics.

# 2 Methods

In this section the device fabrication, the nanoprobing technique, the developed capacitance model and the geometric effects are discussed. **Figure 1** describes the device fabrication and the nanoprobing technique in detail, which will also be discussed in section 2.1 and 2.2.

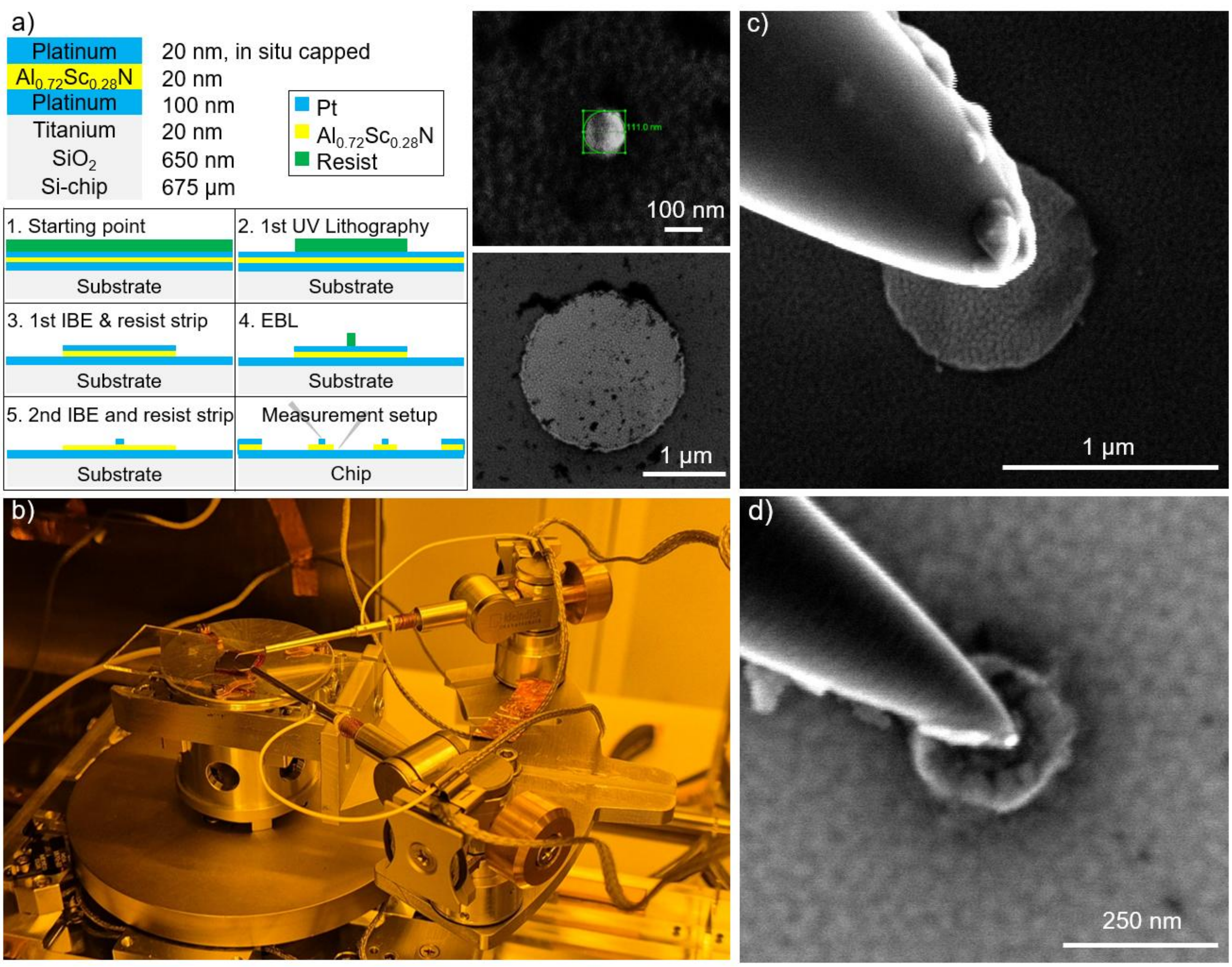

**Figure 1:** Device structure, fabrication, and in-situ electrical nanoprobing. **a)** Schematic of the $Al_{0.72}Sc_{0.28}N$ capacitor stack and the fabrication process flow combining UV lithography and electron beam lithography (EBL) to define devices from 2 µm down to 100 nm in diameter. Inset SEM micrographs show the representative top-down views of the resulting Pt/AlScN/Pt devices. **b)** Photograph of the kleindiek MM3A-EM micromanipulators used for direct electrical probing inside the SEM chamber. **c), d)** High-magnification SEM micrographs showing tungsten probe tips making direct, stable contact with **c)** a 1 µm in diameter and **d)** a 250 nm diameter device. **d)** Demonstrating the precision of the characterization technique.

## 2.1 Nanocapacitor Array Fabrication

To access intrinsic ferroelectric scaling limits, arrays of ultrathin capacitors on Si substrates utilizing a highly textured Pt bottom electrode stack optimized for c-axis nitride growth [1, 9]. Prior to deposition, substrates underwent a surface cleaning sequence, finalized by an ex-situ $Ar/O_2$ plasma cleaning step to ensure a residue-free interface. The functional 20 nm $Al_{0.72}Sc_{0.28}N$ film was deposited via pulsed DC reactive co-sputtering at 450 °C, utilizing parameters previously established to yield high quality wurtzite phase ferroelectricity [1, 10].

Critically, a 20 nm Pt top electrode was deposited immediately after nitride growth without breaking vacuum (**Figure 1a**). This in-situ capping strategy is key for stabilizing the ferroelectric phase in ultrathin films by preventing surface oxidation and nitrogen loss [10].

Device structuring employed two complementary lithography strategies on separate dies to bridge the micro- and nanoscales. Microscale capacitors (diameters 50 μm-5 μm) were defined on the first chip by UV lithography and ion beam etching (IBE) stopping at the AlScN surface confirmed via secondary-ion mass spectroscopy (SIMS) through the detection of Sc [10]. Nanoscale devices (diameters 2 μm-100 nm) were processed on a second chip using a hybrid approach: UV lithography and IBE was first used to isolate mesas and access the bottom electrode, followed by electron beam lithography (EBL) and IBE to define the active device areas (**Figure 1a**). This dual-chip approach yielded a dataset spanning over three orders of magnitude in area, enabling the direct correlation of physical scaling laws from the bulk-like regime to the nanoscale. Detailed fabrication parameters for both process flows are provided in the Supplementary Information.

## 2.2 In-Situ Electrical Nanoprobing

To eliminate parasitic capacitances and environmental noise inherent to large-area testing, electrical characterization was performed in situ within a FEI NovaNano FEG-SEM 630 vacuum chamber (< 1×10-5 mbar). The setup utilized two Kleindiek MM3A-EM micromanipulators equipped with specialized low-impedance tungsten probes to directly contact the sub-micrometer device electrodes (**Figure 1b**). This direct-probing approach bypasses the need for larger bond pads, allowing for the characterization of intrinsic device performance down to the 100 nm scale.

Probe positioning was achieved piezoelectric actuation system operating in a stick-slip “coarse” mode for macroscopic movement and a piezo-extension “fine” mode for final contact. The actuators provide a theoretical fine-mode resolution of <1 nm [11], enabling precise landing on nanoscale electrodes under high-magnification SEM observation. Custom-shaped tungsten probes, (Omniprobe AutoProbe and Mesoscope) with tip diameters ranging from 80-200 nm, were mechanically bent by ~40° to optimize movement and landing angles (**Figure 1c, d**). Further specifications are provided in the Supplementary Information.

For ferroelectric characterization, the probes were connected through vacuum feedthroughs to an aixACCT TF Analyzer 3000 for large-signal polarization switching and pulsed DC leakage measurements. Small-signal capacitance and dielectric loss were acquired using an HP 4284A Precision LCR meter. Measurements were conducted using a 1 V AC drive signal at a frequency of 1 MHz, with 0.2 V step size and a medium integration time with a delay of 100 ms. Standard open/short corrections to remove cabling parasitics have been performed.

## 2.3 Capacitance Modeling and De-embedding

To isolate the intrinsic dielectric response the deep sub-micrometer devices ($r < 2\ \mu\mathrm{m}$), a physical de-embedding framework was established to decouple the macroscopic circuit parasitics from the mesoscopic probe-sample interactions. While standard open/short corrections compensate for fixture impedances up to the probe tip, they fail to account for dynamic stray fields generated by the probe needle geometry in the near-field limit [7, 8].

The total measured capacitance $C_{\mathrm{meas}}(r)$ was decomposed into three distinct physical components:

$$C_{\mathrm{meas}}(r) = C_{\mathrm{bulk}}(r) + C_{\mathrm{static}} + C_{\mathrm{dynamic}}(r) \tag{1}$$

Here, $C_{\mathrm{bulk}}$ represents the ideal parallel-plate capacitance of the AlScN film. $C_{\mathrm{static}}$ denotes the static stray capacitance of the macroscopic probe shaft. This component manifests as a constant background offset, independent of the microscopic electrode variations, resulting from far-field coupling of the needle body to the substrate [8].

The third component, $C_{\mathrm{dynamic}}(r)$, accounts for the variable fringing fields at the electrode perimeter and the active probe-cone interaction field. The analytical baseline for this edge capacitance was established using the exact solution for a circular disk derived by Kirchhoff [12], with finite-thickness corrections according to Palmer [13]:

$$C_{\mathrm{edge}}(r) = \varepsilon_0 \varepsilon_r r \left[ \ln\left(\frac{16\pi r}{d}\right) - 1 + \ln\left(1 + \frac{2t}{d}\right) \right] \tag{2}$$

To address the divergence of the stray field at the nanoscale, where the electrode fails to effectively shield the probe tip from the substrate, this baseline was scaled by a radius dependent $K_f(r)$. The geometric factor was modeled using a Screened Power Law, which introduces an electrostatic screening length $L$ to describe the transition from unshielded probe-cone interaction to a shielded parallel-plate regime.

$$K_f(r) = K_{\text{base}} + A\left(\frac{d}{r}\right)^n \exp\left(-\frac{r}{L}\right) \tag{3}$$

## 2.4 Modeling of Dielectric Loss Dilution

To accurately quantify the intrinsic dielectric loss ($\tan\delta$) at the nanoscale, the influence of the parasitic capacitance floor was explicitly modeled. Unlike capacitance, which is additive, the measured loss tangent represents a capacitance-weighted average of the distinct physical contributions. In the deep sub-micrometer regime ($r < 200\ \text{nm}$), the signal is dominated by the parasitic floor, which was modeled as a composite of two distinct physical pathways:

1. Lossless Static Stray ($C_{\text{static}}$): Attributed to the probe arm and manipulator body coupling through the vacuum/air dielectric, modeled as effectively lossless ($\tan\delta_{\text{meas}} \approx 0$).
2. Lossy Dynamic Stray ($C_{\text{dynamic}}$): Attributed to the active tip cone interacting with the ferroelectric AlScN mesa, which carries the material's intrinsic loss ($\tan\delta_{\text{AlScN}}$).

Consequently, the measured loss tangent $\tan\delta_{\text{meas}}$ describes a dilution of the intrinsic material loss by the lossless parasitic background, defined by the mixture equation:

$$\tan\delta_{\text{meas}}(r) \approx \frac{C_{\text{dynamic}}(r)\cdot\tan\delta_{\text{AlScN}} + C_{\text{static}}}{C_{\text{dynamic}}(r) + C_{\text{static}}} \tag{4}$$

This formulation was used to de-embed the geometric "lossless dilution" effect from the radius-dependent loss data, allowing for the extraction of the true bulk conductance scaling described in Section 3.2.

## 2.5 Geometric and Microstructural Scaling Metrics

To distinguish between continuum-level behavior and confinement effects, the microstructural evolution across the scaling range was quantified. The average grain diameter of polycrystalline AlScN film was determined to be 31.15 by the linear intercept method applied to top view scanning electron micrographs [14]. This value aligns with the columnar grain morphology typically reported for sputtered AlScN thin films [1].

Based on this microstructural baseline, the statistical ensemble of grains contributing to the ferroelectric switching was calculated for each electrode size. As illustrated in (Figure S2), the scaling spans five orders of magnitude: macroscopic devices (Ø 50µm) average over a bulk ensemble of $N \approx 2.5 \times 10^6$ grains, whereas the smallest nanodevices (Ø 165 nm) confine the measurement to a discrete set of $N < 30$ grains. Concurrently, the perimeter-to-area ratio ($P/A$) increases inversely with the radius (**Figure S2c**). For the sub-micrometer devices, this ratio increases by two orders of magnitude compared to the macroscopic reference, implying that the measurement boundary conditions are theoretically dominated by edge interactions. Defining these geometric scaling metrics is essential to verify whether the ferroelectric stability observed in Section 3 arises from bulk averaging or persists even within the localized "few-grain" confinement regime.

# 3 Results and Discussion

The scaling of ferroelectric $Al_{0.72}Sc_{0.28}N$ devices from the macroscopic regime (Ø 50 µm) down to the deep sub-micrometer range (Ø 165 nm) enables the investigation of wurtzite crystal stability under significant microstructural confinement. As the electrode dimensions contract, the measurement volume transitions from a bulk ensemble of over $10^6$ grains to a discrete cluster of fewer than 30 grains, concurrently increasing the relative influence of edge boundary conditions and parasitic impedance. Consequently, distinguishing intrinsic ferroelectric scaling from extrinsic measurement artifacts becomes the central challenge of the analysis. In the following sections, these contributions are systematically deconstructed, de-embedding geometric fringing fields and parasitic series resistance, to isolate the intrinsic response of the material. The results demonstrate that the ferroelectric polarization and coercive field of AlScN remain stable throughout this scaling range, exhibiting stable switching behavior even when confined to the regime of polycrystalline granularity.

### 3.1 Probing the Limits of Capacitance: Fringing Fields at the Nanoscale

To extract the intrinsic permittivity of the AlScN film, we first isolated the device capacitance by de-embedding the static parasitic floor determined in Section 2.3. **Figure 2a** displays the raw capacitance scaling over 3 orders of magnitude from macroscopic pads ($r \approx 25\ \mu\text{m}$) down to devices ($r \approx 82\ \text{nm}$). High-resolution SEM images **Figure S2** reveal an extension of the effective area of the Pt top electrode touching the AlScN film. Therefore, an effective radius $r_{eff}$ was utilized to extract the real area as accurately as possible. While the response of the capacitance follows the expected $C \propto r^2$ trend in the bulk regime, a significant deviation is observed at sub-micrometer dimensions.

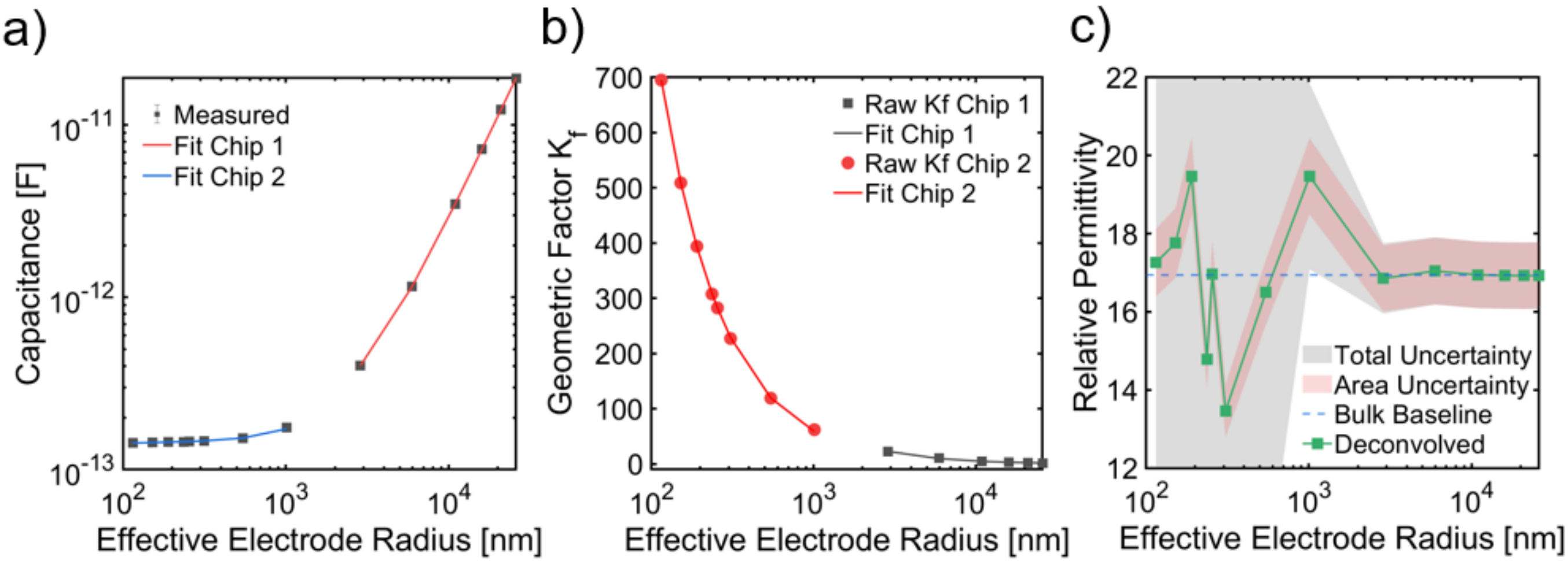

**Figure 2:** Capacitance de-embedding and the Screened Power Law model. **a)** Raw measured small-signal capacitance as a function of the effective electrode radius, demonstrating the severe deviation from ideal parallel-plate scaling at the nanoscale due to probe-sample fringing fields. **b)** The extracted geometric factor ($K_f$) capturing the divergence of the stray field. The exponent transition strictly defines the shift from macroscopic remote-probe coupling to a highly localized, super-linear field concentration at the conductive electrode boundary. **c)** The final deconvolved relative permittivity ($\varepsilon_r$), demonstrating a size-independent intrinsic dielectric response ($\varepsilon_r \approx 17$) across five orders of magnitude in device area. The error bounds represent the root-sum-square of the geometric and instrumental measurement uncertainties.

The extracted geometric factor $K_f$ reveals a significant enhancement in stray capacitance, rising from ≈1 in the micrometer regime to >100 at the nanoscale (**Figure 2b**). This divergence reflects the transition between three distinct physical regimes established in micromechanical (MEMS) and scanning probe microscopy (SPM) literature. First in the plate regime ($r > 10\ \mu\text{m}$), the devices behave as a classical parallel plate capacitor. The geometric factor converges to unity ($K_f \approx 1.0 - 1.5$) confirming that the Kirchhoff-Palmer formulation is sufficient for macroscopic devices. Second in the transition regime ($300\ \text{nm} < r < 5\ \mu\text{m}$), the $K_f$ values ($\approx 10 - 100$) exceed predictions for standard vertical-wall electrodes. We attribute this enhancement to the specific microstructural profile of the top electrode. High-resolution SEM analysis reveals a tapered trapezoidal profile that terminates in a nanoscale "foot" at the perimeter (**Figure S1**). The formation of such a tapered "knife-edge" geometry during IBE is well documented. Cross-sectional FIB-SEM analysis of comparable PT/AlScN/Pt devices has explicitly visualized the formation of these sharp metallic fins at the electrode perimeter [15]. Electrostatically, this sharp terminus acts as a conductive wedge with an acute opening angle ($\beta \to 0°$). According to the analytical solution for potential singularities at conducting edges, the electric field divergence at such a "knife-edge" scales as $E \propto r^{-0.5}$, significantly stronger than the $E \propto r^{-0.33}$ divergence of a standard 90° vertical wall [16]. This geometric field concentration effectively expands the active electrical area beyond the lithographic electrode footprint. This enhancement is further compounded by 3D corner singularities arising from the polygonal faceting of the etched electrodes [17]. Third, in the tip/cone regime ($r < 300\ \text{nm}$), where the sub-micrometer limit is hit, the "parallel plate" assumption physically breaks down as the electrode area vanishes. The system effectively transitions to a "Probe-Sample" interaction dominated by the tip geometry [7, 18]. In this regime, the theoretical bulk capacitance drops to the attofarad range ($C_{\text{bulk}} < 0.5\ \text{fF}$). Experimental characterization of the zero-radius limit ($r \rightarrow 0$) established a stable total parasitic floor of ($\approx 142\ \text{fF}$), which aligns with measurement data of small devices depicted in **Figure 2a**. After subtracting the static arm contribution ($C_{\text{static}} \approx 70\ \text{fF}$), the remaining dynamic stray capacitance, originating from the active tip and cone region, persists at a saturated level of $\approx 72\ \text{fF}$. Consequently, the ratio $K_f \approx C_{\text{dynamic}}/C_{\text{device}}$ mathematically diverges ($K_f > 100$) to reconcile the constant dynamic stray field with the vanishing electrode area. This divergence is not a fitting artifact but a necessary mathematical consequence of maintaining a continuous capacitance model across the transition from plate-dominated to tip-dominated physics.

By applying the Screened Power Law model (Eq. 3) to capture these effects, we successfully extracted a size-independent relative permittivity of $\varepsilon_r \approx 17 \pm 3$ across the entire scaling range (**Figure 2c**), which aligns with previously reported AlScN permittivity [19, 20]. The error bars in **Figure 2c** incorporate both the dimensional uncertainty ($\sigma_{\text{area}} = 5\%$) the instrumental accuracy of the LCR meter and the statistical error (see Supplementary Note 7).

The physical validity of the Screened Power Law is confirmed by the consistency of the extracted fitting parameters across the dimensional transition. Table 1 summarizes the coefficients obtained from the radius-dependent fitting of the geometric factor $K_f$ for both the macroscopic wafer-prober setup (Chip 1) and the nanoprober setup (Chip 2).

**Table 1:** Extracted Fitting Parameters for the Screened Power Law Model. A fixed Intrinsic $\varepsilon_r \approx 16.94$ extracted from the larger pads of Chip 1. The actual thickness of the AlScN film was estimated at ≈17 nm to align with the permittivity previously reported [19, 20].

| Parameter | Symbol | Chip 1 | Chip 2 | Physical Interpretation |
|---|---|---|---|---|
| Fit Quality | $R^2$ | 0.99981 | 0.99995 | Goodness of fit. |
| Static stray Capacitance (locked) [F] | $C_{\text{static}}$ | 8.0e-14 | 7.0e-14 | Far-field probe stray field coupling. |
| Base Factor | $K_{\text{base}}$ | 1.00 | 1.50 | Recovery of the Kirchhoff limit. Chip 1 converges to the ideal parallel plate (1.0); Chip 2 shows edge enhancement (1.5) attributed to sloped electrode sidewalls. |
| Coupling Amplitude | $A$ | 2107 | 5942 | Magnitude of the probe-sample stray field coupling. |
| Scaling Exponent | $n$ | 0.86 | 1.12 | Field divergence rate. The shift from sub-linear ($n < 1$) to super-linear ($n > 1$) marks the onset of the Conductive Fin singularity. |

The extracted Base Factor ($K_{\text{base}}$) provides a crucial boundary condition check. For the macroscopic devices (Chip 1), $K_{\text{base}}$ converges to the lower bound of 1.00, confirming that in the limit of large radii ($r \rightarrow \infty$), the devices behave as ideal parallel-plate capacitors with negligible parasitic overhead. For the nanodevices (Chip 2), the baseline reaches the upper bound at 1.50. This enhancement is attributed to the trapezoidal cross-section of the EBL-defined electrodes. Unlike the ideal cylinder assumed by Kirchhoff, the nanodevices exhibit sloped sidewalls (verified by SEM, Figure S2), which increases the effective surface area available for linear fringing fields, resulting in a constant geometric offset.

Most significantly, the Scaling Exponent ($n$) reveals the fundamental change in the electrostatic boundary conditions. In the macroscopic regime, the exponent $n \approx 0.86$ indicates a sub-linear divergence, consistent with the remote coupling of a probe shaft held at a 45° angle. However, in the nanoscale regime, the exponent transitions to $n \approx 1.12$. This super-linear divergence confirms that the field concentration is no longer governed by standard edge effects but is driven by the "Conductive Fin" singularity ($E \propto r^{-0.5}$) discussed in Supplementary Note 5. The probe tip, now effectively larger than the device itself, couples aggressively to the AlScN mesa and the tapered "knife-edge" of the electrode foot, creating the intense energy density concentrations that necessitate the Screened Power Law correction.

## 3.2 Deconstructing Dielectric Loss in Nanoscale AlScN

Accurately resolving the intrinsic dielectric loss ($\tan\delta$) of nanoscale ferroelectrics is fundamentally challenged by the dominance of extrinsic impedance contributions. As the device radius shrinks, the measurement signal becomes increasingly susceptible to two distinct parasitic mechanisms: resistive losses in the ultra-thin electrodes and capacitive shunting by the vacuum environment [7, 21]. To isolate the material response, we implemented a physical two-stage de-embedding sequence addressing the resistive and capacitive parasitic contributions separately.

Standard open/short corrections effectively remove cabling inductance but fail to account for spreading resistance ($R_{\text{spread}}$) inherent to the point-contact geometry of the probing setup. In the limit of thin-film electrodes ($t \ll r$), the current injection from the tungsten tip is not uniform, creating a distributed resistance profile that manifests as an artificial increase in dielectric loss at high frequencies.

Rather than employing an arbitrary empirical fit, we modeled this parasitic contribution using the analytical formalism for spreading resistance in circular contacts [21]. By incorporating the specific resistivity of the textured Pt films ($\rho \approx 10.6 \times 10^{-8} \Omega\text{m}$ [22]) and the tip-contact geometry, we calculated a radius-dependent resistance baseline (see Supplementary Note 6). As shown in **Figure 3a**, this physical model (blue triangles) predicts the series resistance at ≈ 6-8 Ω across the device range.

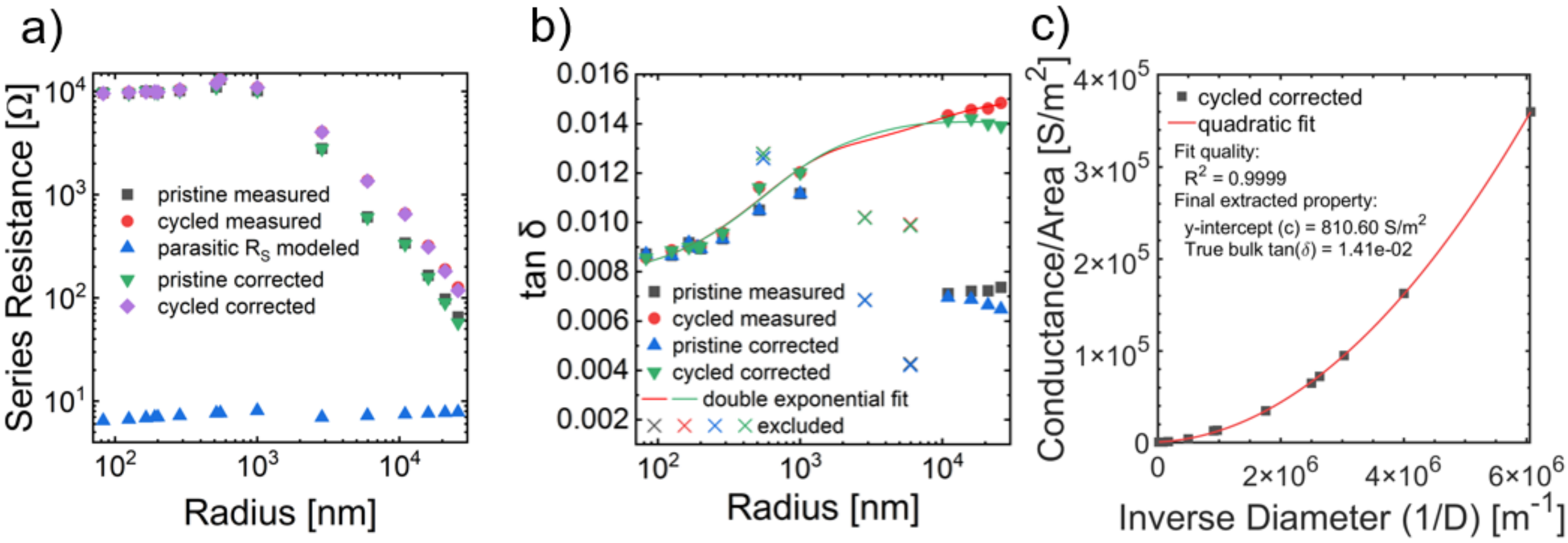

**Figure 3:** De-embedding parasitic impedance and intrinsic dielectric loss extraction. **a)** Series resistance ($R_s$) as a function of device radius. The physical spreading resistance model (blue triangles) isolates the current-constriction at the nanoscale probe contact, while the measurement plateaus at $10^4$ Ω due to the static vacuum impedance limit. **b)** Measured dielectric loss tangent ($tan\ \delta$) for pristine and cycled devices. The apparent loss suppression in the deep sub-micrometer regime is governed by a geometric dilution artifact, accurately modeled by a double-exponential decay representing the lossless vacuum shunt of the probe shaft. **c)** Conductance scaling per unit area ($G/A$) versus inverse diameter ($1/D$). The quadratic fit mathematically extracts the size-independent bulk conductance density, yielding a true intrinsic $tan\ \delta \approx\ 1.41\ \%$ for the fully cycled ferroelectric state.

To de-embed this effect, the raw parallel impedance measured by the LCR meter ($C_p, R_p$) was transformed into its series equivalent ($Z_s$). The modeled parasitic resistance was then subtracted from the total real part of the impedance ($R_{\text{s,int}} = R_{\text{s,meas}} - R_{\text{model}}$), effectively placing the reference plane at the intrinsic capacitor interface.

Additionally, **Figure 3a** displays the resulting dielectric series resistance as a function of device radius. The data reveals two distinct scaling regimes. First, in the macroscopic regime ($r > 1\ \mu\text{m}$), the resistance follows the expected inverse-area power law ($R \propto r^{-2}$), indicated by the linear slope on the log-log plot. In this range, the device impedance is dominated by the AlScN film capacitance, and the subtraction of the electrode spreading resistance (<10 Ω) provides a minor but necessary correction to maintain linearity at largest radii. Second, in the nanoscale plateau ($r < 1\ \mu\mathrm{m}$), the resistance deviates from the power law and saturates at a constant plateau of ≈$10^4$ Ω. This plateau represents the impedance resolution limit of the setup. In the nanoscale regime, the measured signal is no longer dominated by the shrinking device, but by the constant parasitic capacitance floor ($C_{\mathrm{total}} \approx 142\ \mathrm{fF}$) and its associated loss.

Following the resistive correction, the dielectric loss tangent was analyzed (**Figure *3*b**). Contrary to the degradation typically associated with sidewall damage, the measured $\tan\delta$ decreases from ≈ 1.41 % in the macroscopic devices to a floor of ≈ 0.76 % at the nanoscale. This behavior is identified as a geometric dilution effect. In the deep sub-micrometer regime, the vanishing device capacitance ($C_{\mathrm{device}} < 1\ \mathrm{fF}$) becomes negligible to the "lossless" static stray capacitance ($C_{\mathrm{static}} \approx 70\ \mathrm{fF}$) coupling through vacuum, which effectively shunts the ferroelectric signal and dilutes the measured $\tan\delta$.

Crucially, however, the loss does not drop to zero but saturates at a parasitic baseline of ≈0.76%. This residual loss arises because a significant fraction of the stray field ($C_{\mathrm{dynamic}}$) still couples through the AlScN mesa. Unlike the vacuum parasitics, these fringing fields interact with lossy ferroelectric lattice, preventing total signal dilution. The measured response at the nanoscale is therefore a weighted average of the intrinsic device loss, the lossless vacuum shunt, and the mesa-coupled stray loss.

This interpretation is quantitatively supported by the capacitance extraction from Section 3.1. The measured loss tangent in the nanoscale limit is physically determined by the ratio of lossy mesa coupling ($C_{\mathrm{dynamic}}$) to the total parasitic capacitance ($C_{\mathrm{total}} = C_{\mathrm{static}} + C_{\mathrm{dynamic}}$). Substituting the extracted values ($C_{\mathrm{static}} \approx 70\ \mathrm{fF}$, $C_{\mathrm{dynamic}} \approx 72\ \mathrm{fF}$) and the intrinsic bulk loss ($\tan\delta_{bulk} \approx 1.41\%$) into the dilution equation:

$$\tan\delta_{floor} \approx \frac{C_{\mathrm{dynamic}}}{C_{\mathrm{static}}+C_{\mathrm{dynamic}}} \cdot \tan\delta_{bulk} \tag{5}$$

yields a theoretical floor of ≈0.71 %. This prediction is in excellent agreement with the experimental saturation level (≈0.76 %), confirming that the observed drop in $\tan\delta$ is not an artifact of material degradation, but a predictable geometric consequence of the "lossless" vacuum-coupled probe arm diluting the signal.

To quantify the magnitude of this suppression, the radius-dependent transition was modeled using a phenomenological double-exponential decay function:

$$\tan\delta(r) = y_0 + A_1 e^{\frac{r}{t_1}} + A_2 e^{\frac{r}{t_2}} \tag{6}$$

The high quality of the fit ($R^2 > 0.991$) confirms that the observed drop is a systematic geometric effect rather than random measurement noise. The extracted asymptotic limit yields a bulk loss of $y_0 \approx\ 1.41\%$ consistent with the intrinsic material properties. Significantly, the negative amplitude terms ($A_1 \approx\ -0.002$, $A_2 \approx -0.004$) reveal that in the sub-micrometer limit, the parasitic dilution suppresses the apparent dielectric loss by approximately 45 %.

Since both the capacitance and the measured loss tangent saturate (as shown in **Figure 2a** and **3b**), the derived series resistance ($ESR \approx \tan\delta/\omega C$) mathematically converges to a constant value. This confirms that direct $R_s$ or $\tan\delta$ extraction is invalid in the sub-micrometer limit, necessitating the conductance-based de-embedding.

To recover the true material loss without relying on empirical fitting parameters, we analyzed the conductance per unit area ($G/A$) as a function of inverse diameter ($1/D$) (**Figure 3c**). Following standard area-perimeter decomposition techniques, the total conductance density scales as [21]:

$$\frac{G_{total}}{A} = \sigma_{\mathrm{bulk}} + \frac{\sigma_{\mathrm{edge}}}{D} + \frac{\sigma_{\mathrm{const}}}{D^2} \tag{7}$$

Here, the slope terms quantify three distinct physical mechanisms. First, the $\sigma_{\mathrm{bulk}}$ y-intercept, representing the intrinsic area-dependent conductance of the ferroelectric domains. Second, the linear term ($\sigma_{\mathrm{edge}}/D$), quantifying the perimeter-dependent leakage arising from etch-induced sidewalls and the probe tip. Third, the quadratic term ($\sigma_{\mathrm{const}}/D^2$), which models the constant background conductance that does not scale with device size. This term captures strictly localized effects (such as corner leakage) as well as the fixed loss associated with the far-field stray coupling to the AlScN film. The quadratic fit ($R^2 > 0.9999$) yields an intercept of $\approx 810\ \mathrm{S/m^2}$. Converting this conductance back to a loss tangent yields an intrinsic $\tan\delta_{\mathrm{bulk}} \approx 1.41\%$. This value is in excellent agreement with macroscopic reference measurements, confirming that the damping mechanisms of the ferroelectric domain walls are preserved even when confined to a polycrystalline cluster of fewer than 30 grains.

Finally, contrasting the pristine and cycled datasets reveals the physical origin of the extracted bulk loss. As shown in **Figure 3b**, the pristine devices (grey squares) exhibit a significantly lower baseline loss ($\tan\delta_{\mathrm{pristine}} \approx 0.7\%$) compared to the cycled state $\tan\delta_{\mathrm{cycled}} \approx 1.41\%$). This systematic offset is consistent with the ferroelectric wake-up effect observed in polycrystalline AlScN [1, 23].

In the as-deposited state, domain walls are largely pinned by local defect complexes (primarily nitrogen vacancies), suppressing their contribution to the dielectric susceptibility. Upon electric field cycling, these domains are mechanically unpinned [23], allowing them to vibrate reversibly in response to the AC drive signal. The observed increase in $\tan\delta$ therefore does not indicate material degradation, but rather quantifies the extrinsic domain wall damping contribution to the total loss [24]. Consequently, the extracted value of $\approx\ 1.41\%$ represents the true operational loss figure for the active ferroelectric memory state.

## 3.3 Ferroelectricity at the Nanoscale

Validating ferroelectric switching dynamics typically relies on small-signal capacitance-voltage (C-V) measurements, which manifest as characteristic "butterfly" loops. While the qualitative shape of these loops remains surprisingly robust in raw normalized data down to the nanoscale, extracting true quantitative metrics presents a severe metrological challenge. In the deep sub-micrometer limit ($r\ <\ 1\ \mu\mathrm{m}$), the intrinsic device capacitance drops into the attofarad range. To overcome signal-to-noise ratio constraints and at this scale, signal averaging, Fast Fourier Transformation (FFT) low pass filters and multi-peak fitting was employed. Multiple sequential C-V sweeps were acquired for each device, with the numbers of averages from 2 sweeps up to 9; see Supplementary Note 8. The initial sweep of every sequence was systematically excluded from the averages. This protocol decouples the transient ferroelectric wake-up effect, where initial domain unpinning causes a shift in the baseline, ensuring that only the stabilized, steady-state hysteresis is analyzed [1, 23].

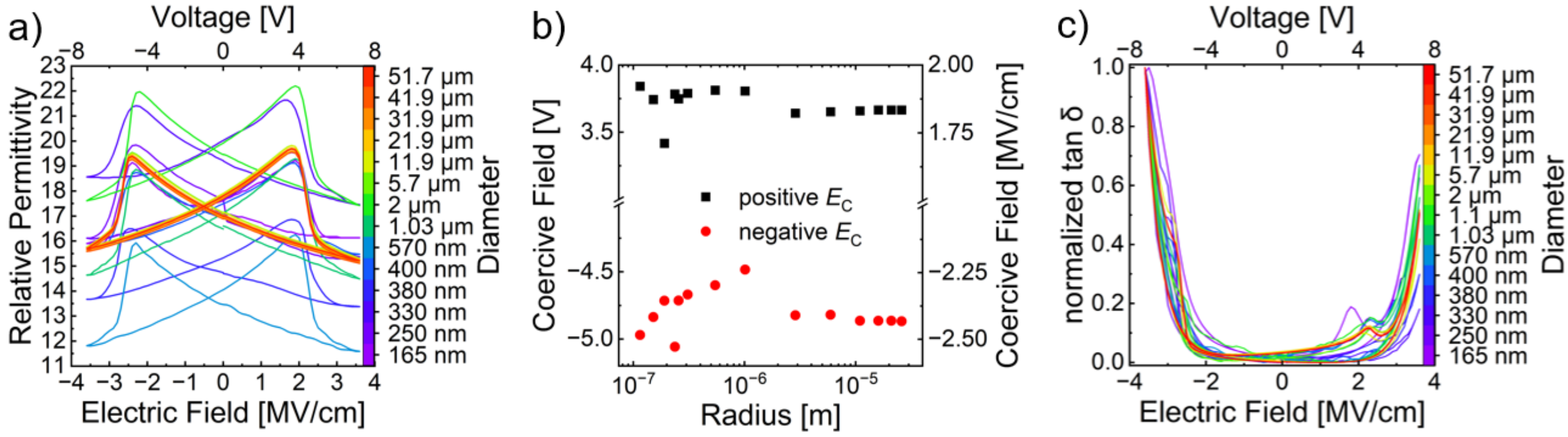

**Figure 4:** Small-signal ferroelectric switching and localized perimeter leakage. **a)** De-embedded relative permittivity vs. electric field (C-V) loops across the dimensional spectrum. **b)** Extracted coercive field ($E_C$) scaling. The systematic hardening of the negative $E_C$ at sub-micrometer dimensions is driven by intense negative field emission at the top-electrode taper. This localized leakage acts as a parallel shunt, inducing a series voltage drop that forces the analyzer to apply a higher external bias to achieve polarization reversal. c) Normalized dynamic loss tangent vs. electric field. The amplitude variance near $E_C$ strictly reflects few-grain nucleation statistics at the nanoscale, while the universal high-field U-shape divergence marks the transition to direct Fowler-Nordheim tunneling [25, 26], defining the absolute physical limit of the LCR measurement principle.

While the de-embedding framework successfully collapses five orders of magnitude of area-dependent capacitance into a narrow intrinsic permittivity band ( $\varepsilon_r \approx 12 - 22$), a minor residual variance between the individual loops is observed (**Figure 4a**). This spread is not a limitation of the model, but rather reflects real-world physical boundaries. At the nanoscale, slight lithographic variations in the electrode edge profile introduce minor fluctuations in the local fringing efficiency. Furthermore, as the measurement volume confines to fewer than 30 discrete grains, the local dielectric response naturally begins to deviate from the macroscopic continuum average due to localized strain and crystallographic variance.

A minor, systematic offset in the absolute coercive field ($E_{\mathrm{C}}$) values is observed at the dimensional crossover the microscale devices (Chip 1) and the nanoscale devices (Chip 2) (**Figure 4b**). This shift is attributed to the devices originating from two distinct deposition runs, which included a variation in the top Pt electrode thickness (30 nm for Chip 1 versus 20 nm for Chip 2). In ultrathin wurtzite ferroelectrics, the coercive field is highly sensitive to the exact active film thickness, interfacial depolarization fields, and residual clamping stress from the electrode boundaries [20]. Consequently, these batch-to-batch structural variations naturally manifest as a minor perturbation in the measured $E_{\mathrm{C}}$.

Beyond this batch offset, a measurement-dependent scaling trend of $E_{\mathrm{C}}$ emerges after the dimensional crossover. Under quasi-static C-V conditions (**Figure 4b**), the magnitude of the negative $E_{\mathrm{C}}$ progressively increases at sub-micrometer dimensions, increasing from approximately -4.5 MV/cm to -5.0 MV/cm. This shift is driven by the increasing dominance of the perimeter-to-area ratio in conjunction with the specific measurement protocol. As established in the dry etching of III-nitride semiconductors, the energetic ion bombardment during the IBE process preferentially sputters lighter nitrogen atoms. This physical damage inherently generates a non-stoichiometric, nitrogen-deficient boundary layer rich in deep $V_{\mathrm{N}}$ defect states along the exposed electrode sidewalls [27]. During the C-V sequence, the applied voltage initializes with a negative-first sweep utilizing a 100 ms delay per step. With the top Pt electrode acting as the cathode, the tapered electrode perimeter concentrates the electric field, driving localized field emission directly into this damaged perimeter region. The resulting prolonged charge injection manifests as a pronounced negative leakage divergence in the dynamic loss profiles (**Figure 4c**). Electrostatically, this highly localized perimeter leakage acts as a parallel resistive shunt during polarization reversal. As the leakage current increases, it induces a corresponding voltage drop across the series resistance of the measurement setup. Consequently, the effective electric field penetrating the ferroelectric bulk is suppressed relative to the applied terminal voltage. To achieve the intrinsic critical field necessary for domain nucleation, the analyzer must apply a higher external voltage, yielding the apparent $E_{\mathrm{C}}$ hardening observed in the highly scaled devices.

Finally, the dynamic dielectric loss was evaluated (**Figure 4c**). While the static de-embedding framework successfully isolates the absolute bulk loss at zero bias, resolving the absolute dynamic loss tangent under a sweeping high-field bias at the nanoscale exceeds the phase-angle resolution limits of conventional LCR meters. As the intrinsic device capacitance drops below 1 fF, the dynamic change in the loss angle driven by domain wall motion becomes entirely eclipsed by the constant 70 fF vacuum stray capacitance. Consequently, the isolated dynamic loss signatures are presented as normalized values to track the relative domain wall kinetics.

Within these normalized profiles, distinct local maxima emerge near the coercive fields (≈ ± 2.0 MV/cm), representing the extrinsic damping contribution of domain walls during polarization reversal. However, as device dimensions scale into the deep sub-micrometer regime, a pronounced device-to-device variance in the amplitude and precise field-position of these dynamic loss peaks is observed. Crucially, because the top Pt electrode was strictly in-situ capped without vacuum break, this variance cannot be attributed to interfacial molecular contamination or inadequate chemical cleaning. Instead, it reflects the fundamental stochastic reality of extreme microstructural confinement. At 165 nm in diameter, the measurement volume encompasses fewer than 30 discrete grains. In this limit, the macroscopic continuum average physically breaks down; minor statistical variations in localized crystallographic orientation, intrinsic grain-boundary density, and exact sidewall etch-morphology govern the specific nucleation kinetics of each individual capacitor, naturally manifesting as variance in the dynamic loss signatures.

Furthermore, as the applied field exceeds ±3 MV/cm, the loss tangent diverges exponentially to form a pronounced U-shape across the entire dimensional spectrum, spanning from the macroscopic 51.7 µm capacitors down to the 165 nm devices (**Figure 4c**). This universal divergence is an extrinsic artifact of the LCR measurement principle. At these extreme biases, the ultrathin 20 nm AlScN film intrinsically enters a severe field-driven leakage regime (e.g., Fowler-Nordheim tunneling) [25, 26], independent of the device perimeter. In the standard parallel-circuit model utilized by LCR meters, this exponential collapse of the parallel device resistance ($R_\mathrm{p}$) mathematically inflates the calculated loss tangent ($\tan\delta \propto R_\mathrm{p}^{-1}$), completely overwhelming the fundamental AC response. This severe, area-independent leakage masking explicitly defines the absolute boundary of conventional high-field small-signal metrology for ultrathin wurtzites, necessitating a transition to the large-signal leakage-compensation techniques established in Section 3.4 to accurately recover the intrinsic polarization hysteresis.

## 3.4 J-E loops

To validate the scalability of the ferroelectric switching mechanism, we performed a systematic area-dependent PUND study (**Figure 5**). The primary challenge in the sub-µm² regime is the diminishing signal-to-noise ratio (SNR) of the displacement current, which necessitates advanced signal processing to extract accurate switching kinetics.

For devices with active areas below Ø5 µm, the raw current response is frequently obscured by instrumentation noise and background leakage (**Figure S7**). **Figure 5a** illustrates the signal processing workflow applied to a representative device. The "raw" trace (grey) exhibits significant high-frequency noise components. To isolate the ferroelectric switching current, first the fitted leakage has been subtracted, a Fast Fourier Transform (FFT) low pass filter was applied (red trace), effectively attenuating noise while preserving the switching transient. A subsequent fit (blue trace) was utilized to extract the precise coercive field, ensuring data fidelity at the detection limit.

**Figure 5b** displays the evolution of Current Density vs. Electric Field (J-E) loops across a diameter range spanning two orders of magnitude, from 50 µm to 190 nm. The overlapping current density up until Ø 1µm peaks demonstrate a remarkable stability of the ferroelectric switching mechanism. Crucially, clear switching peaks are resolvable even at the 190 nm and 570 nm diameter scales, confirming that ferroelectricity remains accessible in wurtzite-type AlScN is maintained in the deep sub-µm² regime.

The extraction of the coercive field $E_\mathrm{C}$ as a function of the active area is plotted in **Figure 5c**. While the negative coercive field (red circles) remains relatively stable around -4.9 to -5.0 MV/cm, a distinct scaling trend is observed for the positive coercive field (black squares). As the active area decreases from $10^{-13}$ m² to $10^{-9}$ m², the positive $E_\mathrm{C}$ increases from ~4.6 MV/cm to nearly ~5.0 MV/cm. This asymmetry and hardening of the coercive field in highly scaled devices is consistent with the device asymmetry. This asymmetry and hardening of the positive coercive field in highly scaled devices is directly governed by the pronounced leakage asymmetry inherent to the device architecture. As shown in **Figure 5d**, while the negative leakage density remains relatively stable with only a slight increase from ~100 A/cm² to ~200 A/cm², the positive leakage diverges exponentially as the perimeter-to-area ratio increases (**Figure S2c**). During the rapid positive switching pulse, the highly defective bottom Pt/AlScN growth interface acts as the cathode. The resulting injected electron flux preferentially routes through the highly conductive, IBE-damaged sidewalls. Electrostatically, this severe positive perimeter leakage acts as a parallel resistive shunt. As this leakage current diverges, it induces a significant localized voltage drop across the series resistance of the measurement setup, physically suppressing the effective electric field penetrating the ferroelectric bulk. To overcome this extrinsic voltage loss and achieve the intrinsic critical field necessary for polarization reversal, the external analyzer must apply a substantially higher voltage. This area-dependent leakage masking strictly dictates the pronounced positive $E_\mathrm{C}$ hardening observed at the nanoscale.

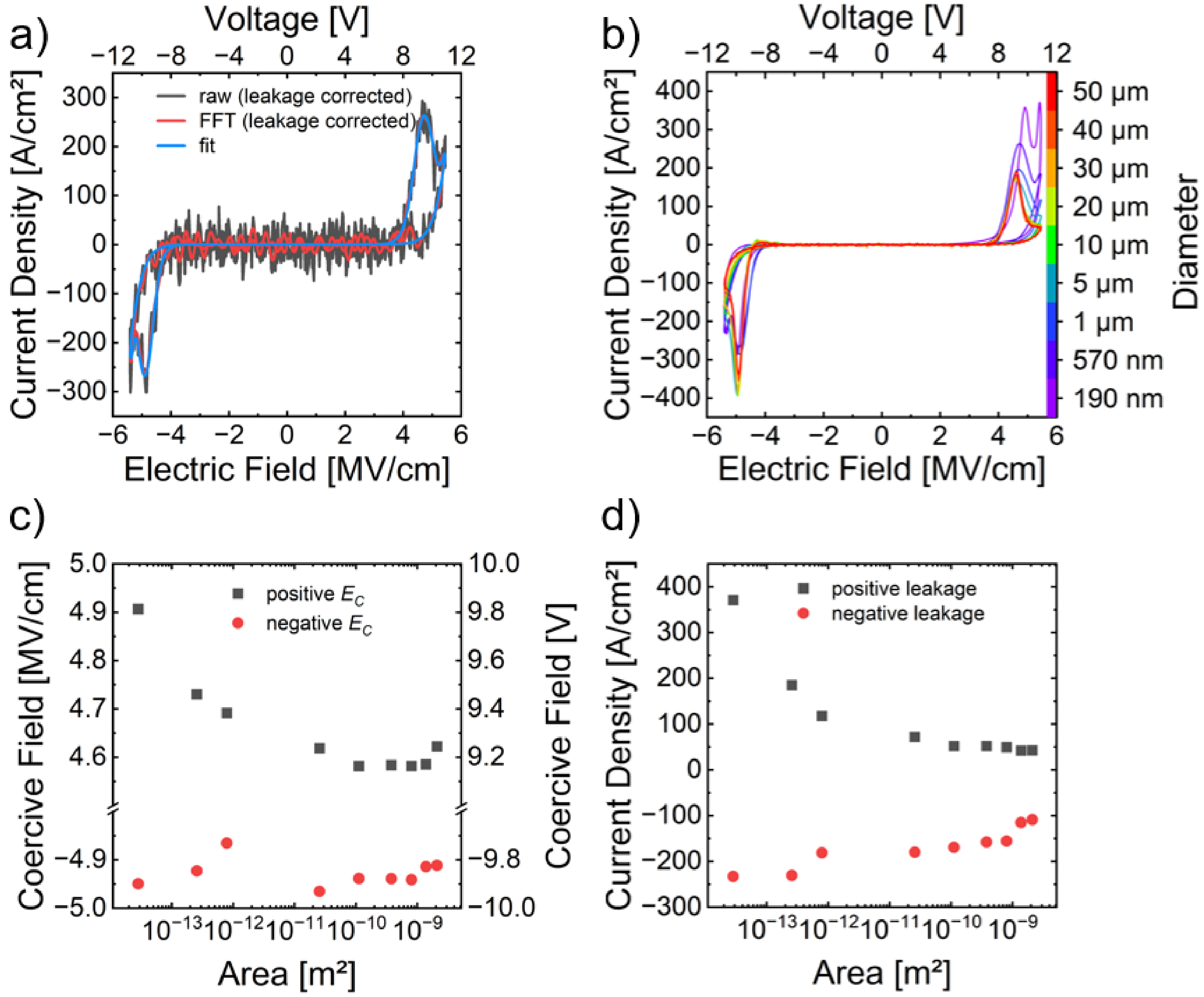

**Figure 5:** Large-signal J-E characterization and history-dependent kinetics at 10 kHz. **a)** Signal processing methodology for the sub-micrometer limit. The raw leakage-corrected current (grey) is processed via a Fast Fourier Transform (FFT) low-pass filter (red) and multi-peak fitting (blue) to recover the pristine switching transient from the attofarad noise floor. **b)** Overlay of J-E loops for device diameters ranging from 50 µm down to 190 nm, demonstrating robust switching across all scales. **c)** Extracted $E_C$ stability versus area $10^{-13}$ m² to $10^{-9}$ m² (Ø190 nm-Ø50 µm), scaling under fast dynamic driving (PUND). The severe hardening of the positive $E_C$ directly correlates with the leakage asymmetry shown in d). **d)** Leakage current density vs. area. The exponential divergence of the positive leakage, driven by bottom-interface injection routing through the conductive sidewalls, acts as a massive dynamic shunt, severely suppressing the internal positive electric field and dictating the observed $E_C$ hardening.

# 4 Conclusion

We have demonstrated a quantitative, in-situ nanoprobing framework for the electrical characterization of deep sub-micrometer ferroelectric capacitors, successfully bypassing the massive parasitic limitations of lithographic bond pads. By analytically modeling the electrostatic singularities of the near-field probe-sample interaction, we decoupled the intrinsic attofarad-level device response from the dominating femtofarad-level geometric parasitics. This de-embedding protocol proves that the apparent suppression of dielectric loss at the nanoscale is fundamentally a geometric dilution artifact, confirming that 20 nm thick AlScN films retain intrinsic, bulk-like ferroelectric switching down to 165 nm in diameter.

Ultimately, this framework provides the analytical toolset required to engineer the interface-dominated physics of future high-density wurtzite ferroelectrics. By explicitly isolating the impact of perimeter-driven leakage shunts from intrinsic bulk domain kinetics, this scaling study serves as a critical diagnostic tool for optimizing sub-micrometer fabrication. Specifically, quantifying the exponential leakage divergence at the ion-beam-etched sidewalls underscores the fundamental necessity for advanced post-etch recovery protocols. As established in III-nitride high electron mobility transistor (HEMT) processing, integrating wet-chemical damage removal (e.g., using TMAH or KOH-based etchants) with conformal passivation (e.g., PECVD $SiN_x$) is mandatory to mitigate uncompensated edge states [28, 29]. Translating these specific passivation strategies to nanoscale ferroelectrics will be essential to suppress extrinsic edge leakage, prevent localized voltage-masking effects, and preserve pristine polarization switching at the ultimate scaling limit.

## 4.1 References

1 Fichtner, S., Wolff, N., Lofink, F., Kienle, L., Wagner, B. (2019) AlScN: A III-V semiconductor based ferroelectric. *J. Appl. Phys.*, **125** (11).

2 Schönweger, G., Dasenbrook, D., Kyoushi, N., Guido, R., Petraru, A., Mikolajick, T., Schröder, U., Kohlstedt, H., Fichtner, S. (2026) Nitride Ferroelectric Domain Wall Memory for Next-Generation Computing. *Adv Elect Materials*, **12** (3), e00616.

3 Balke, N., Maksymovych, P., Jesse, S., Herklotz, A., Tselev, A., Eom, C.-B., Kravchenko, I.I., Yu, P., Kalinin, S.V. (2015) Differentiating Ferroelectric and Nonferroelectric Electromechanical Effects with Scanning Probe Microscopy. *ACS Nano*, **9** (6), 6484–6492.

4 Balke, N., Maksymovych, P., Jesse, S., Kravchenko, I.I., Li, Q., Kalinin, S.V. (2014) Exploring local electrostatic effects with scanning probe microscopy: implications for piezoresponse force microscopy and triboelectricity. *ACS Nano*, **8** (10), 10229–10236.

5 Vasudevan, R.K., Balke, N., Maksymovych, P., Jesse, S., Kalinin, S.V. (2017) Ferroelectric or non-ferroelectric: Why so many materials exhibit “ferroelectricity” on the nanoscale. *Appl. Phys. Rev.*, **4** (2).

6 Gruverman, A., Alexe, M., Meier, D. (2019) Piezoresponse force microscopy and nanoferroic phenomena. *Nat Commun*, **10** (1), 1661.

7 Fumagalli, L., Ferrari, G., Sampietro, M., Casuso, I., Martínez, E., Samitier, J., Gomila, G. (2006) Nanoscale capacitance imaging with attofarad resolution using ac current sensing atomic force microscopy. *Nanotechnology*, **17** (18), 4581–4587.

8 Gomila, G., Toset, J., Fumagalli, L. (2008) Nanoscale capacitance microscopy of thin dielectric films. *Journal of Applied Physics*, **104** (2).

9 Su, J., Fichtner, S., Ghori, M.Z., Wolff, N., Islam, M.R., Lotnyk, A., Kaden, D., Niekiel, F., Kienle, L., Wagner, B., Lofink, F. (2022) Growth of Highly c-Axis Oriented AlScN Films on Commercial Substrates. *Micromachines*, **13** (5), 783.

10 Schönweger, G., Petraru, A., Islam, M.R., Wolff, N., Haas, B., Hammud, A., Koch, C., Kienle, L., Kohlstedt, H., Fichtner, S. (2022) From Fully Strained to Relaxed: Epitaxial Ferroelectric Al 1-x Sc x N for III-N Technology. *Advanced Functional Materials*, **32** (21).

11 Shi, C., Luu, D.K., Yang, Q., Liu, J., Chen, J., Ru, C., Xie, S., Luo, J., Ge, J., Sun, Y. (2016) Recent advances in nanorobotic manipulation inside scanning electron microscopes. *Microsyst Nanoeng*, **2** (1), 16024.

12 Kirchhoff, G.R. (1877) Zur Theorie des Condensators, 144–162.

13 Palmer, H.B. (1937) The Capacitance of a Parallel-Plate Capacitor by the Schwartz-Christoffel Transformation. *Trans. Am. Inst. Electr. Eng.*, **56** (3), 363–366.

14 Test Methods for Determining Average Grain Size, E04 Committee, West Conshohocken, PA.

15 Lu, H., Schönweger, G., Wolff, N., Ding, Z., Petraru, A., Streicher, I., Kohlstedt, H., Kübel, C., Leone, S., Kienle, L., Fichtner, S., Gruverman, A. (2025) Al 1-x Sc x N-Based Ferroelectric Domain-Wall Memristors. *Advanced Functional Materials*, **35** (46), 2503143.

16 Jackson, J.D. (2009) *Classical electrodynamics*, 3rd edn, Wiley, Hoboken, NY.

17 Yuan, C.P. and Trick, T.N. (1982) A simple formula for the estimation of the capacitance of two-dimensional interconnects in VLSI circuits. *IEEE Electron Device Lett.*, **3** (12), 391–393.

18 S. Hudlet, M. Saint Jean, C. Guthmann, J. Berger (1998) Evaluation of the capacitive force between an atomic force microscopy tip and a metallic surface. *The European Physical Journal B - Condensed Matter and Complex Systems*.

19 Fichtner, S., Wolff, N., Krishnamurthy, G., Petraru, A., Bohse, S., Lofink, F., Chemnitz, S., Kohlstedt, H., Kienle, L., Wagner, B. (2017) Identifying and overcoming the interface originating c-axis instability in highly Sc enhanced AlN for piezoelectric micro-electromechanical systems. *J. Appl. Phys.*, **122** (3).
20 Schönweger, G., Islam, M.R., Wolff, N., Petraru, A., Kienle, L., Kohlstedt, H., Fichtner, S. (2023) Ultrathin Al 1− x Sc x N for Low-Voltage-Driven Ferroelectric-Based Devices. *physica status solidi (RRL) – Rapid Research Letters*, **17** (1), 2200312.
21 Schroder, D.K. (2005) *Semiconductor Material and Device Characterization*, Wiley.
22 Dutta, S., Sankaran, K., Moors, K., Pourtois, G., van Elshocht, S., Bömmels, J., Vandervorst, W., Tőkei, Z., Adelmann, C. (2017) Thickness dependence of the resistivity of platinum-group metal thin films. *Journal of Applied Physics*, **122** (2).
23 Gremmel, M. and Fichtner, S. (2024) The interplay between imprint, wake-up, and domains in ferroelectric Al0.70Sc0.30N. *Journal of Applied Physics*, **135** (20).
24 Damjanovic, D. (1998) Ferroelectric, dielectric and piezoelectric properties of ferroelectric thin films and ceramics. *Rep. Prog. Phys.*, **61** (9), 1267–1324.
25 Jabbari, I., Baira, M., Maaref, H., Mghaieth, R. (2020) Evidence of Poole-frenkel and Fowler-Nordheim tunneling transport mechanisms in leakage current of (Pd/Au)/Al0.22Ga0.78N/GaN heterostructures. *Solid State Communications*, **314-315**, 113920.
26 Wang, R., Zhou, J., Zheng, S., Zhu, F., Sun, W., Xu, H., Li, B., Liu, Y., Hao, Y., Han, G. (2025) Cryogenic Ferroelectric Behavior of Wurtzite Ferroelectrics.
27 Pearton, S.J., Shul, R.J., Ren, F. (2000) A Review of Dry Etching of GaN and Related Materials. *Materials Research Society Internet Journal of Nitride Semiconductor Research*, **5** (1), e11.
28 Green, B.M., Chu, K.K., Chumbes, E.M., Smart, J.A., Shealy, J.R., Eastman, L.F. (2000) The effect of surface passivation on the microwave characteristics of undoped AlGaN/GaN HEMTs. *IEEE Electron Device Lett.*, **21** (6), 268–270.
29 Vetury, R., Zhang, N.Q., Keller, S., Mishra, U.K. (2001) The impact of surface states on the DC and RF characteristics of AlGaN/GaN HFETs. *IEEE Trans. Electron Devices*, **48** (3), 560–566.

# Supplementary Information

## Supplementary Note 1: Detailed Fabrication Parameters

**Substrate Preparation:** AlScN nanocapacitors were fabricated on 1×1 cm² Si chips (675 µm thickness) coated with 650 nm of thermally grown $SiO_2$. The bottom electrode stack consisted of a 20 nm Ti adhesion layer and a 100 nm Pt seed layer. Prior to deposition, chips were cleaned sequentially ultrasonic bath in acetone, isopropanol, and deionized water, followed by an ex-situ plasma cleaning step (2.5 sccm Ar / 5 sccm $O_2$, 250 W, 5 min) to remove organic residues.

**Thin Film Deposition:** The functional layers were deposited using an Oerlikon (now Evatec) CLUSTERLINE 200 multi-source sputtering system. The 20 nm $Al_{0.72}Sc_{0.28}N$ film was grown by pulsed DC reactive co-sputtering at a substrate temperature of 450 °C. Process parameters were controlled at 680 W for the Al target and 320 W for the Sc target, with gas flows of 15 sccm $N_2$ and 7.5 sccm Ar. The 20 nm Pt top electrode was deposited immediately after the nitride growth without breaking vacuum to prevent interface oxidation. The Pt top electrode was deposited at 450 °C and 400W with a 80 sccm gas flow of Krypton.

**Lithography and Etching:** Fabrication was split across two processing distinct flows to optimize pattern fidelity for different size regimes:

**1. Microscale Fabrication (Chip 1, 50 µm – 5 µm):** Devices in the 50 µm to 5 µm range were patterned using standard UV lithography (Karl Süss MA6, AZ 1518, MicroChemicals GmbH). The top Pt electrode was structured via Ion Beam Etching (IBE), timed to stop with a controlled over-etch at the AlScN surface verified by secondary-ion mass spectroscopy (SIMS) through the detection of Sc, leaving the ferroelectric layer continuous across the die.

**2. Nanoscale Fabrication (Chip 2, 2 µm – 100 nm):** Deep-sub micrometer devices required a two-step hybrid process to ensure electrical isolation and precise definition:

- **Step A (Mesa Definition):** Large-area mesas (20×20µm²) were defined using UV lithography and IBE. In this step, the etch proceeded through the AlScN layer and stopped at the bottom Pt electrode, creating isolated islands and exposing the bottom contact.

**Step B (Device Definition):** Nanoscale active areas were defined on top of the mesas using an electron beam lithography FEI NovaNano SEM FEG-SEM 630 equipped with a Raith ELPHY Quantum/Plus EBL system. A negative resist (AR-N 7520.18, Allresist GmbH), an acceleration voltage of 20 kV and with doses varying from 0.01µC to 1000 µC was used. A second IBE step transferred this pattern into the top Pt electrode, stopping at the AlScN surface with a controlled over-etch.

## Supplementary Note 2: Nanoprobing Setup and Probe Characterization

**Actuation and Control:** The probes were manipulated using Kleindiek MM3A-EM piezoelectric actuators. Coarse and fine positioning were controlled via a gamepad interface mapped to the manipulator axes (X, Y, Z). The system operates on a piezoelectric stick-slip principle for coarse motion (step size ~40 nm-1µm in linear axis) and stator deflection for fine motion. The theoretical resolution in fine mode is approximately 0.02 nm per digital step for the linear axis and 0.3 nm for rotational axes [1].

**Probe Specifications:** To ensure stable electrical contact on varying device scales, three types of tungsten probes were utilized:

1. **High-Resolution Probes:** Mesoscope probes were selected for the smallest devices and to measure the small-signal capacitance and dielectric loss measurement series in the nanoprober. They feature a shaft diameter ranging from 80 nm to 100 nm (measured at a height of 50 nm above the apex).
2. **Robust Probes:** For larger devices and general contact, Omniprobe AutoProbe tips were used, characterized by a tip radius of 0.5 µm (corresponding to a diameter of ~200 nm at 50 nm above the apex) and a taper angle of 10°-13°.

3. **Waferprober Probes:** AmericanProbes tungsten wire probes on a nickel shank were utilized with a tip radius of 0.35 µm.

All probes used in the nanoprober setup were inserted into adapters and mechanically bent downwards by approximately 40° at a distance of 2-3 mm from the tip. This geometrical adaptation was essential to align the probe's actuation axes with the sample's coordinate system, ensuring that control inputs resulted in orthogonal X, Y and Z translation.

## Supplementary Note 3: Area Calculation via SEM Segmentation

Accurate extraction of the intrinsic permittivity and coercive fields strictly requires the physical measurement of the effective electrical area $A_{\mathrm{eff}}$, rather than reliance on ideal dimensions. As detailed in **Figure S1**, high-resolution SEM image segmentation was utilized for all devices measured with the SEM Nanoprober.

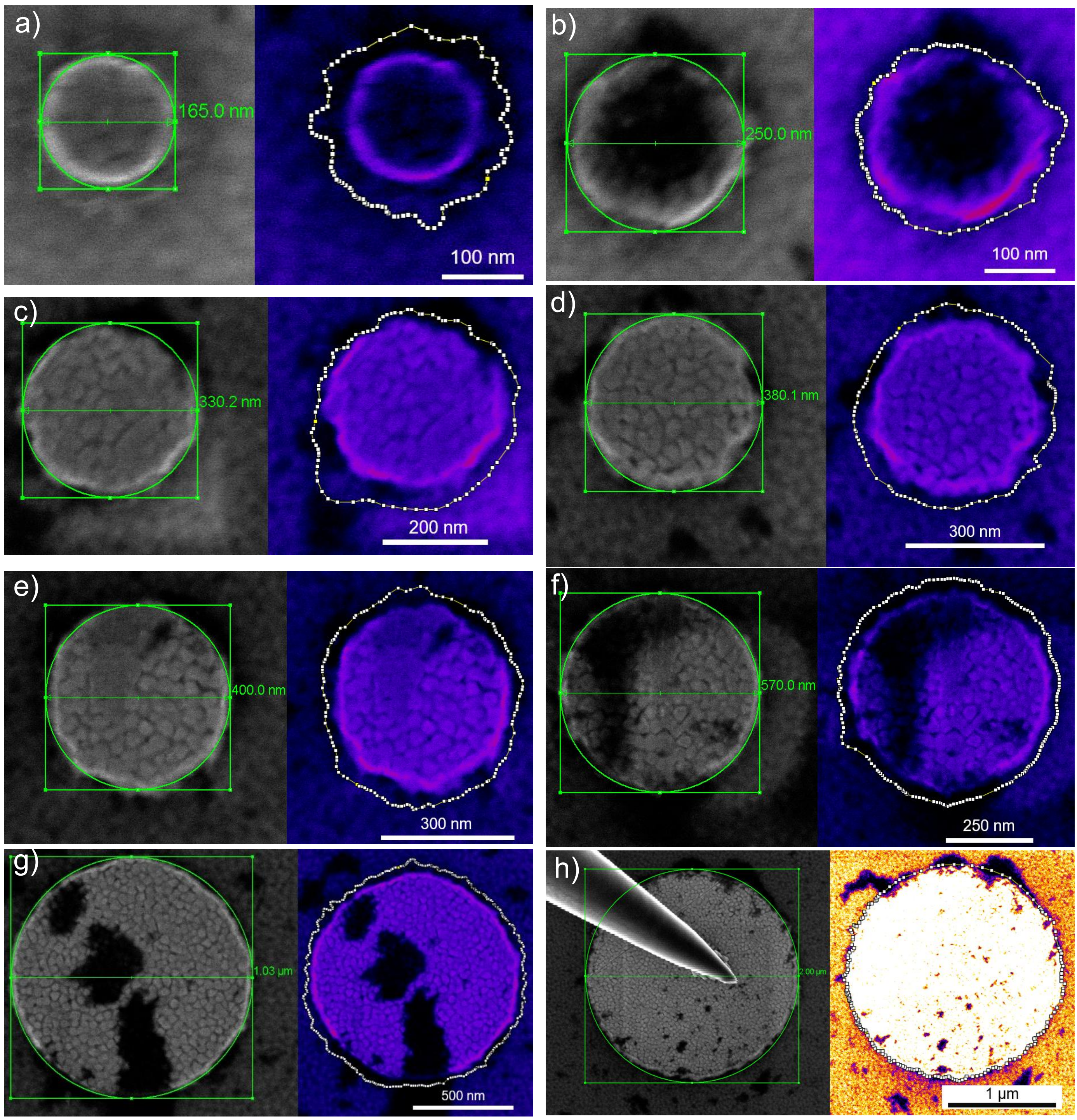

**Figure S1:** High-resolution top-down SEM micrographs of the representative 165 nm, 250 nm, 330 nm, 380 nm, 400 nm, 570 nm, 1.03 µm and 2 µm devices. Due to the tapered top electrode generated during the secondary ion beam etching process, a distinct conductive 'foot' extends outward from the primary electrode cylinder. Thickness threshold processing (ImageJ Fiji) of this structural boundary was utilized to calculate the true effective electrical area ($A_{\text{eff}}$) and effective radius ($r_{\text{eff}}$), establishing the dimensional baseline required for the subsequent capacitance and conductance de-embedding models.

The tapered electrode edge introduced by the second IBE process results in a distinct, conductive geometric 'foot' extending radially from the primary electrode cylinder. Because this highly conductive Pt layer remains in direct contact with the underlying AlScN, it actively participates in the electrostatic fringing and Fowler-Nordheim charge injection mechanisms [2, 3]. Image analysis (ImageJ Fiji) was employed to map the polygonal perimeter of this structural foot. The total enclosed area was then utilized to calculate the effective radius $\left(r_{eff} = \sqrt{A_{eff}/\pi}\right)$, which served as the baseline geometric input for all subsequent analytical capacitance and conductance models. **Table S1** summarizes the extracted values.

**Table S1:** Geometric extraction via SEM segmentation. Comparison of the nominal lithographic dimensions and the true effective electrical geometries utilized for the sub-micrometer devices. The effective radius ($r_{\text{eff}}$) and effective area ($A_{eff}$) were mathematically extracted via polygonal perimeter mapping (ImageJ Fiji) to account for the physical 'foot' generated during the secondary ion beam etching process. These empirical values provide the rigorous boundary conditions required for the analytical de-embedding models.

| Device Diameter | Radius [nm] | Effective radius [nm] | Area [m²] | Effective Area [m²] |
|---|---|---|---|---|
| 2 µm | 1000 | 991 | $3{,}14159\times10^{-12}$ | $3.08478\times10^{-12}$ |
| 1.03 µm | 515 | 545 | $8{,}33229\times10^{-13}$ | $9{,}31419\times10^{-13}$ |
| 570 nm | 285 | 320 | $2{,}55176\times10^{-13}$ | $3{,}11040\times10^{-13}$ |
| 400 nm | 200 | 232 | $1{,}25664\times10^{-13}$ | $1{,}68786\times10^{-13}$ |
| 380 nm | 190 | 214 | $1{,}13411\times10^{-13}$ | $1{,}44094\times10^{-13}$ |
| 330 nm | 165 | 190 | $8{,}55299\times10^{-14}$ | $1{,}14030\times10^{-13}$ |
| 250 nm | 125 | 135 | $4{,}90874\times10^{-14}$ | $5{,}68080\times10^{-14}$ |
| 165 nm | 82,5 | 115 | $2{,}13825\times10^{-14}$ | $4{,}14780\times10^{-14}$ |

## Supplementary Note 4: Geometric and Microstructural Scaling Metrics

To mathematically distinguish between bulk-continuum physics and finite-size confinement effects, the specific microstructural boundary conditions of the AlScN capacitors were quantified. The mean grain diameter of the wurtzite AlScN film was determined to be 31.15 nm via the standard ASTM E112 linear intercept method applied to high-resolution top-view SEM micrographs [4]. Based on this microstructural baseline, the statistical ensemble of grains ($N$) actively contributing to the ferroelectric response was calculated as a function of the electrode area (**Figure S2a**). At the macroscopic scale (50 µm), the electrode encompasses $N > 10^6$ grains, ensuring that localized crystallographic variations perfectly average into a uniform continuum response. However, as the device diameter scales below 200 nm, $N$ drops below 30. In this discrete confinement regime, stochastic variations in individual grain orientation and defect density strictly dictate the macroscopic electrical readout. Concurrently, the perimeter-to-area ($P/A$) ratio diverges (**Figure S2b**). While the theoretical geometric limit for a perfect cylinder dictates a strict $2/r$ divergence, the empirical data extracted via SEM segmentation heavily deviates from this idealization. As plotted in **Figure S2c**, the physical nanodevices exhibit a systematically higher fringing efficiency. This structural enhancement is the direct physical consequence of the polygonal faceting induced by the electron-beam lithography and the conductive geometric 'foot' remaining after the ion beam etching process. This higher-order geometric divergence definitively establishes the physical scaling threshold where edge boundary conditions electrostatically overpower the intrinsic bulk volume.

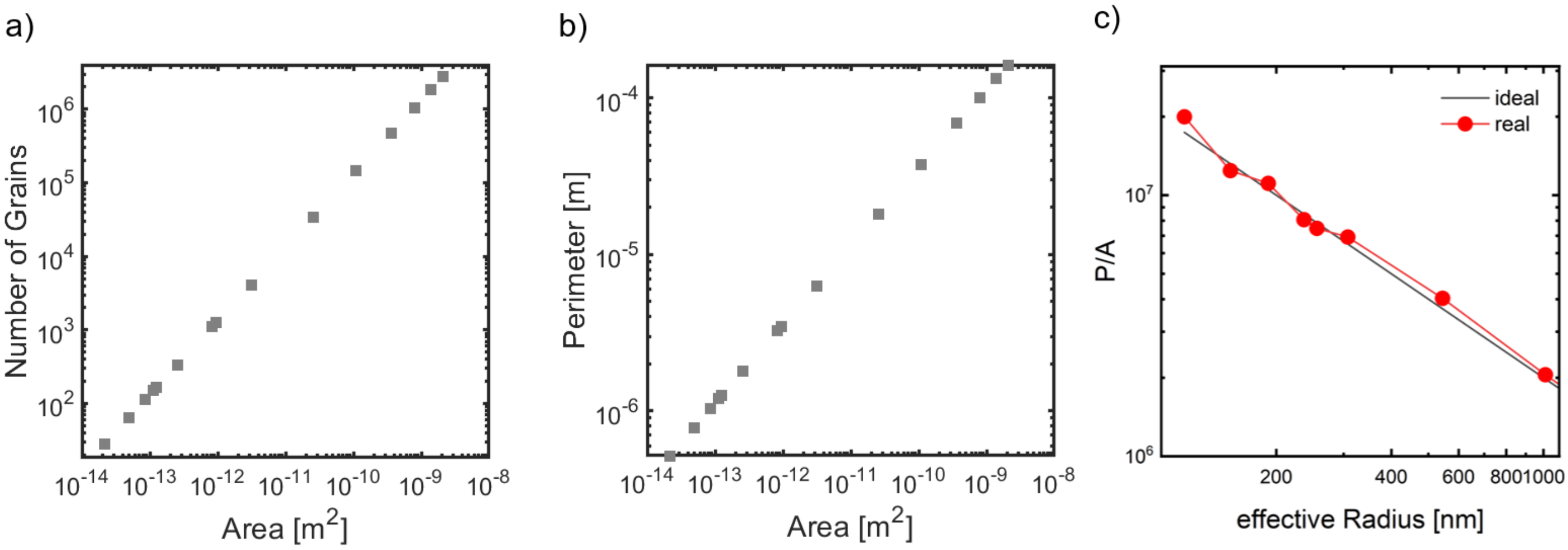

**Figure S2:** Geometric and microstructural scaling metrics. **a)** Calculated statistical ensemble of AlScN grains contributing to the ferroelectric response as a function of electrode area. The grain count ($N$) was derived using a mean grain diameter of 31.15 nm (determined via ASTM E112 line intercept method), revealing a transition from a bulk continuum ($N > 10^6$) in macroscopic devices to a discrete confinement regime ($N < 30$) in the smallest nanodevices. **b)** Scaling of the electrode perimeter versus area, highlighting the geometric constraints where edge boundary conditions become increasingly dominant relative to the bulk film area. c) Extracted perimeter-to-area ($P/A$) ratio representing the fringing efficiency. The strict $2/r$ divergence dictates the theoretical scaling limit for an ideal cylinder. The measured structural data (red circles) explicitly deviates from this idealization due to the EBL-induced polygonal faceting and the conductive electrode foot, confirming an enhanced geometric fringing efficiency that overpowers the intrinsic bulk volume at the nanoscale.

## Supplementary Note 5: Physical Capacitance Model and Geometric De-embedding

**Analytical Edge Capacitance Baseline:** The intrinsic fringing field at the capacitor perimeter was modeled using the analytical solution for a circular parallel-plate capacitor. While simplified perimeter approximations ($C \propto P$) are often employed in macroscopic modeling, they do not capture the field curvature effects inherent in sub-micrometer devices, as they neglect the logarithmic field curvature terms ($\propto \ln(r/d)$) derived for circular discs [5]. The baseline edge capacitance $C_{\text{edge}}(r)$, was therefore calculated using the circular disk formulation derived by Kirchhoff [5]:

$$C_{\text{edge}}(r) = \varepsilon_0 \varepsilon_r r \left[\ln\left(\frac{16\pi r}{d}\right) - 1 + \ln\left(1 + \frac{2t}{d}\right)\right] \tag{1}$$

where $\varepsilon_0$ is the vacuum permittivity, $\varepsilon_r$ is the relative permittivity of the AlScN film, and $d$ is the dielectric thickness. The final logarithmic term accounts for the finite top-electrode thickness $t$ as derived by Palmer [6]. As shown in comparative studies using numerical boundary element methods (BEM) [7], this analytical solution provides an accurate description of the edge field for capacitors with low aspect ratios ($d/r \ll 1$), validating its use as the physical baseline for AlScN thin films ($d/r \approx 0.1 - 0.6$) prior to the application of probe-specific geometric corrections.

**Static Stray Capacitance and Environmental Effects:** The static stray capacitance was estimated by the asymptotic behavior of the measured capacitance limit as $r \to 0$.

- **Wafer Prober Setup (Chip 1):** For macroscopic devices on a continuous AlScN film a residual capacitance of $C_{\text{static}} \approx 80$ fF was extracted. In this configuration, the probe needle is held at a steep angle (≈45°), but the stray fields couple through the high-permittivity AlScN film ($\varepsilon_r \approx 17$).
- **Nanoprober Setup (Chip 2):** For nanoscale devices on isolated mesas, a static stray capacitance of $C_{\text{static}} \approx 70$ fF was determined. Although the probe arm is positioned in close proximity (parallel) to the substrate, enhancing the geometric coupling, the removal of AlScN film around mesa ($\varepsilon_{enc} \approx 1$) suppresses the dielectric contribution. These competing effects result in a comparable static offset despite the distinct experimental geometries.

**Screened Power Law for Nanoscale Probes:** In the near-field limit, the active probe cone (the bottom 10-20 μm) couples to the substrate, creating a dynamic stray capacitance $C_{\text{dynamic}}$. This interaction is modeled by scaling the analytical baseline with a geometric factor $Kf(r)$, defined by a screened power law:

$$C_{\text{dynamic}}(r) = K_f(r) \cdot C_{\text{edge}}(r) \tag{2}$$

$$K_f(r) = K_{\text{base}} + A\left(\frac{d}{r}\right)^n \exp\left(-\frac{r}{L}\right) \tag{3}$$

where $K_{\text{base}}$ recovers the standard edge fringing (≈1-2), A is the amplitude of the near-field coupling, and n describes the field concentration at the tapered electrode foot. The screening length L was estimated at 15 μm, consistent with the effective electrostatic height of the active probe taper reported in scanning probe microscopy literature [8].

**Derivation of the Conductive Fin Singularity:** The enhanced geometric factor ($K_f$) observed in the Transition regime ($300\text{ nm} < r < 5\text{ μm}$) is attributed to the "Conductive Fin" effect, where electrode perimeter terminated in a nanoscale wedge rather than a vertical wall. We model this utilizing the electrostatic solution for potential singularity at the conducting edge derived by Jackson [9]. The electric field strength $E$ near a conducting corner varies with distance $r$ from the tip according to:

$$E(r) \propto r^{v-1} \tag{4}$$

where

$$v = \frac{\pi}{2\pi - \beta} \tag{5}$$

Here, $\beta$ is the opening angle of the conductor.

- Case A: Standard vertical Sidewall ($\beta = 90° = \pi/2$). The exponent becomes $v = \pi/1.5\pi = 2/3$. The field diverges as $E \propto r^{-0.33}$.
- Case B: Conductive Fin/Knife edge ($\beta \to 0°$). For a tapered foot (width ≈20-30 nm, thickness ≈1 nm), the angle approaches zero. The exponent becomes $v = \pi/2\pi = 1/2$. The field diverges as $E \propto r^{-0.5}$.

Since the stored energy density $u$ in the capacitor scales with the square of the field ($u = \frac{1}{2}\epsilon E^2$), the energy density near the fin scales as $r^{-1.0}$, compared to $r^{-0.66}$ for a vertical wall. This significantly stronger singularity integrates to a higher total fringing capacitance, effectively expanding the active electrical area beyond the lithographic footprint.

**Dielectric Loss Dilution Calculation:** To quantify the magnitude of the apparent dielectric loss suppression at the nanoscale, the radius-dependent measured loss tangent was modeled using a phenomenological double-exponential decay function (Equation 6 in the main text):

$$\tan\delta(r) = y_0 + A_1 e^{\frac{r}{t_1}} + A_2 e^{\frac{r}{t_2}} \tag{6}$$

This function captures the geometric transition from the bulk-dominated measurement regime at macroscopic radii to the parasitic-dominated regime in the deep sub-micrometer limit. The fitting routine ($R^2 > 0.991$) extracted the following parameters:

- $y_0 \approx 0.0141$
- $A_1 \approx -0.00205$
- $A_2 \approx -0.00441$

The macroscopic boundary condition is validated by the asymptotic limit at large device radii ($r \to \infty$). Here, the exponential decay terms vanish, and the measurement converges strictly to $y_0 \approx 1.41\,\%$, accurately reflecting the intrinsic bulk loss of the fully cycled ferroelectric film.

Conversely, the absolute magnitude of the parasitic dilution is evaluated at the ultimate nanoscale boundary condition. As the device radius approaches zero ($r \to 0$), the exponential terms converge to unity. The theoretical zero-radius loss tangent limit is therefore mathematically defined by the sum of the coefficients:

$$\tan\delta(r) = y_0 + A_1 + A_2 \tag{7}$$

$$\tan\delta(r) = 0.0141 - 0.00205 - 0.00441 = 0.00764 \tag{8}$$

This yields a theoretical minimum measured loss of $0.76\ \%$, which aligns with the experimental nanoscale plateau of $0.71\ \%$,. To determine the relative signal suppression induced by this geometric dilution artifact, the maximum parasitic deviation ($A_1 + A_2$) is normalized against the intrinsic bulk baseline ($y_0$):

$$\text{Dilution Suppression} = \frac{|A_1+A_2|}{y_0} = \frac{|-0.00646|}{0.0141} \approx 45.8\ \% \tag{9}$$

This mathematical extraction explicitly confirms the claim in the main text: the 'lossless' vacuum-coupled stray field of the nanoprober dilutes the true material signal, artificially suppressing the apparent dielectric loss by approximately $45\ \%$ prior to the application of the conductance scaling de-embedding protocol.

## Supplementary Note 6: Electrode Resistance Correction

To definitively rule out measurement artifacts originating from the point-contact probing geometry, the parasitic series resistance was analytically modeled and evaluated against the total measured impedance. Unlike standard lithographic probing where a massive contact pad ensures equipotential distribution, the in-situ nanoprobing setup injects current directly from a localized tungsten tip into the ultra-thin Pt electrodes. This geometry inherently induces a radial spreading resistance. The total parasitic series resistance ($R_{\text{model}}$) was modeled as the sum of the radial current spreading within both the top ($R_{\text{TE}}$) and bottom ($R_{\text{BE}}$) electrodes alongside a nominal system contact resistance.

**Analytical Spreading Model:**

Current injected from the probe tip into the center of the top electrode flows radially outward to the device perimeter. In the thin-film limit, this spreading resistance is calculated by integrating the 2D radial sheet resistance from the effective tip contact radius ($r_{\text{tip}}$) to the capacitor radius ($r_{\text{cap}}$):

$$R_{\text{TE}} = \int_{r_{\text{tip}}}^{r_{\text{cap}}} \frac{\rho_{\text{Pt}}}{2\pi r t_{\text{TE}}} = \frac{\rho_{\text{Pt}}}{2\pi r t_{\text{TE}}} \ln\left(\frac{r_{\text{cap}}}{r_{\text{tip}}}\right) \tag{10}$$

where $\rho_{\text{Pt}} \approx 10.6 \times 10^{-8}\Omega\text{m}$ is the established resistivity of sputtered platinum $[10]$. The boundary conditions were strictly defined for the specific dimensional regimes: top electrode thickness $t_{\text{TE}} = 30\ \text{nm}$ and $r_{\text{tip}} \approx 175\ \text{nm}$ for microscale devices (Chip 1); $t_{\text{TE}} = 20\ \text{nm}$ and $r_{\text{tip}} \approx 50\ \text{nm}$ for nanoscale devices (Chip 2).

Similarly, the return current spreading laterally through the 100 nm thick bottom electrode ($t_{\text{BE}}$) from the active device perimeter ($r_{\text{cap}}$) out to an asymptotic effective ground boundary ($L_{eff} = 30\ \mu m$) was modeled as:

$$R_{\text{BE}} = \int_{r_{\text{tip}}}^{r_{\text{cap}}} \frac{\rho_{\text{Pt}}}{2\pi r t_{\text{TE}}} = \frac{\rho_{\text{Pt}}}{2\pi r t_{\text{TE}}} \ln\left(\frac{L_{eff}}{r_{\text{cap}}}\right) \tag{11}$$

**Physical Negligibility of Series Parasitics:**

The total parasitic baseline was established by summing these spatial current paths with the experimentally measured probe-to-electrode static contact resistance ($R_{\text{contact}} \approx 5.0\ \Omega$):

$$R_{\text{model}} = R_{\text{TE}} + R_{\text{BE}} + R_{\text{contact}} \tag{12}$$

As demonstrated in **Figure 3**a, this comprehensive physical model restricts the total parasitic series resistance ($R_{\text{model}}$) to a narrow band between 5 Ω and 8 Ω across the entire dimensional spectrum. In contrast, the total measured equivalent series resistance ($R_{\text{s,total}} = R_{\text{P}}/(1+Q^2)$) scales from approximately 60 Ω at the microscale to an instrumental saturation plateau of $10^4$Ω in the deep sub-micrometer limit. Because the parasitic probe resistance is several orders of magnitude smaller than the total measured equivalent series resistance, its contribution to the overall impedance profile is physically negligible. This strict mathematical de-embedding definitively proves that localized contact-spreading resistance does not cause the observed high-frequency dielectric loss degradation at the nanoscale. Consequently, the apparent loss suppression must be fundamentally governed by parallel vacuum-shunt mechanisms, corroborating the geometric dilution framework established in Supplementary Note 5.

## Supplementary Note 7: Error Analysis and Instrument Accuracy

The uncertainty in the extracted relative permittivity ($\varepsilon_{\mathrm{r}}$) and intrinsic device capacitance is governed by the propagation of error from three independent sources: geometric variance, systematic instrumental limitations, and statistical measurement noise. The total measurement uncertainty was determined using a rigorous Root-Sum-Square (RSS) propagation framework.

**1. Geometric Uncertainty (($\Delta A$):**

The dimensional uncertainty is dictated by the spatial resolution limits of the scanning electron microscopy (SEM) utilized to extract the effective electrode radius ($r_{\mathrm{eff}}$). Due to finite pixel resolution and the polygonal faceting inherent to electron-beam lithography at the deep sub-micrometer scale, a radial uncertainty of $\pm 5$ % was assigned. Because capacitance scales with the square of the radius ($r^2$), this geometric constraint translates to a fixed 10% baseline uncertainty in the active device area ($\Delta A / A \approx 0.1$).

**2. Systematic Instrumental Accuracy ($\Delta C_{\mathrm{sys}}$):**

In the sub-micrometer limit, the total measured parallel impedance $\left(|Z_{\mathrm{m}}| = \left[(1/R_{\mathrm{p}})^2 + (\omega C_{\mathrm{p}})^2\right]^{-1/2}\right)$ increases exponentially, severely degrading the performance of standard bridge-balance metrology. The systematic capacitance error ($\Delta C_{\mathrm{sys}}$) was explicitly quantified using the instrumental accuracy model for the HP 4284A Precision LCR meter operating at 1 MHz and 1 V AC drive bias ($V_{\mathrm{s}} = 1000\ \mathrm{mV}$) [11]:

$$Acc_{\mathrm{sys}}(\%) = A_{\mathrm{basic}} + (K_{\mathrm{b}} \cdot K_{\mathrm{bb}} \cdot 100) + K_{\mathrm{d}} \tag{13}$$

where the baseline accuracy is $A_{\mathrm{basic}} = 0.1$ %, the cable length factor is $K_{\mathrm{bb}} = 2$, and the frequency-dependent degradation term is $K_{\mathrm{d}} = 0.0255$ %. The impedance-proportional error term ($K_{\mathrm{b}}$) captures the resolution loss at the attofarad scale, defined as:

$$K_{\mathrm{b}} = |Z_{\mathrm{m}}| \times 10^{-9} \times \left(1 + \frac{70}{V_{\mathrm{s}}}\right) \tag{14}$$

As the device diameter scales below 200 nm, the extreme magnitude of $|Z_{\mathrm{m}}|$ mathematically forces $K_{\mathrm{b}}$ to dominate the systematic error budget, defining the absolute hardware limit of the measurement setup.

**3. Statistical Variance and SNR ($\Delta C_{\mathrm{stat}}$):**

Beyond systematic limits, random thermal and environmental noise ($\Delta C_{\mathrm{stat}}$) masks the intrinsic displacement current at the nanoscale. Initial analysis of the raw, single-sweep traces for the deep sub-micrometer devices revealed a Signal-to-Noise Ratio (SNR) approaching unity. To isolate the fundamental ferroelectric response, sequential sweep averaging was utilized.

Crucially, because the raw measured capacitance ($C_{\mathrm{mean}}$) is overwhelmingly dominated by the static parasitic stray field (> 140 fF), normalizing the statistical variance by the total measured signal artificially dilutes the apparent error. To provide a physically metric, the absolute statistical noise was calculated via the standard error of the mean ($\mathrm{SEM} = \sigma/\sqrt{N}$) in Farads, and subsequently normalized against the mathematically deconvolved intrinsic device capacitance ($C_{\mathrm{intrinsic}}$). Furthermore, because variance is dynamically voltage-dependent, peaking during stochastic domain nucleation at the coercive field and minimizing at full field saturation, this relative statistical error was extracted as a boundary range (minimum to maximum) across the voltage sweep. As detailed in **Table S2**, the applied averaging protocol successfully suppressed the random noise floor. Specifically, for all device dimensions, this sequential averaging mathematically forced the statistical variance below the systematic measurement accuracy (($\Delta C_{\mathrm{stat}} < \Delta C_{\mathrm{sys}}$). Ensuring the uncompensated random noise is entirely eclipsed by the absolute hardware limit, even at the maximum variance boundary near the coercive fields, we definitively prove that the extracted nanoscale switching transients are driven by true physical polarization reversal rather than localized stochastic artifacts.

**4. Total Propagated Uncertainty**

To establish the absolute metrological limits of the extraction framework, the total absolute capacitance error was calculated by adding the systematic instrument limits and the statistical variance in quadrature:

$$\Delta C_{\text{total}} = \sqrt{C_{\text{sys}}^2 + C_{\text{stat}}^2} \tag{15}$$

As visualized in **Figure S3a**, the extracted intrinsic capacitance accurately tracks the physical model, bounded by this combined absolute measurement uncertainty. Finally, this total electrical uncertainty was propagated with the geometric area uncertainty to determine the absolute error bounds for the deconvolved relative permittivity:

$$\Delta\varepsilon_{\text{r,total}} = \varepsilon_{\text{r}}\sqrt{\left(\frac{\Delta C_{\text{total}}}{C_{\text{intrinsic}}}\right)^2 + \left(\frac{\Delta A}{A}\right)^2} \tag{16}$$

As plotted in **Figure S3b**, this error propagation confirms that the intrinsic dielectric response ($\varepsilon_{\text{r}} \approx 17$) remains statistically robust across five orders of magnitude in device area. Despite the exponential divergence of the parasitic impedance penalties in the deep sub-micrometer regime, the de-embedding protocol mathematically secures the intrinsic material parameters within strict, physically defined confidence intervals.

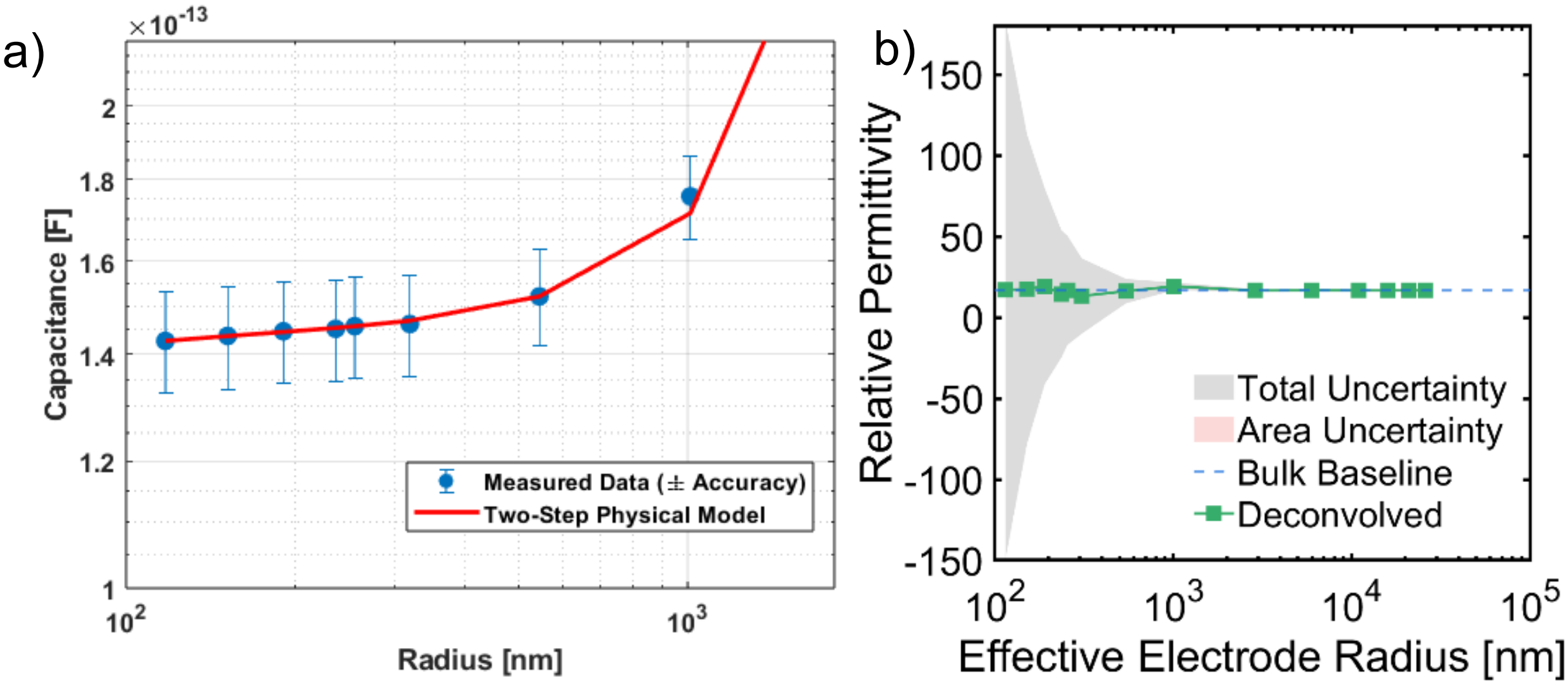

**Figure S3:** Absolute capacitance variance and propagated relative permittivity uncertainty. a) Measured total capacitance versus effective electrode radius (blue circles) overlaid with the combined analytical capacitance model (red line). The error bars represent the absolute systematic instrumental accuracy limit ($\Delta C_{sys}$) defined by the Keysight impedance evaluation framework. b) The fully deconvolved relative permittivity ($\varepsilon_r \approx 17$) plotted across the dimensional scaling range. The shaded regions denote the 5% geometric area uncertainty (red) and the total root-sum-square propagated uncertainty (grey), confirming robust intrinsic material stability despite the exponential divergence of the parasitic measurement error at the deep sub-micrometer limit.

## Supplementary Note 8: CV Data post processing

To extract intrinsic small-signal parameters from deep sub-micrometer devices, raw C-V data was subjected to a systematic post-processing protocol. For each device, up to 9 sequential C-V sweeps were acquired. The initial sweep was systematically excluded to decouple transient ferroelectric wake-up effects and domain unpinning, ensuring only steady-state hysteresis was analyzed. As visualized in **Figure S4**, as the device area shrinks into the sub-micrometer limit, the raw individual cycles become heavily dominated by random instrumental noise. This necessitates sequential averaging to mathematically suppress the statistical variance ($\Delta C_{\text{stat}}$) prior to frequency-domain filtering.

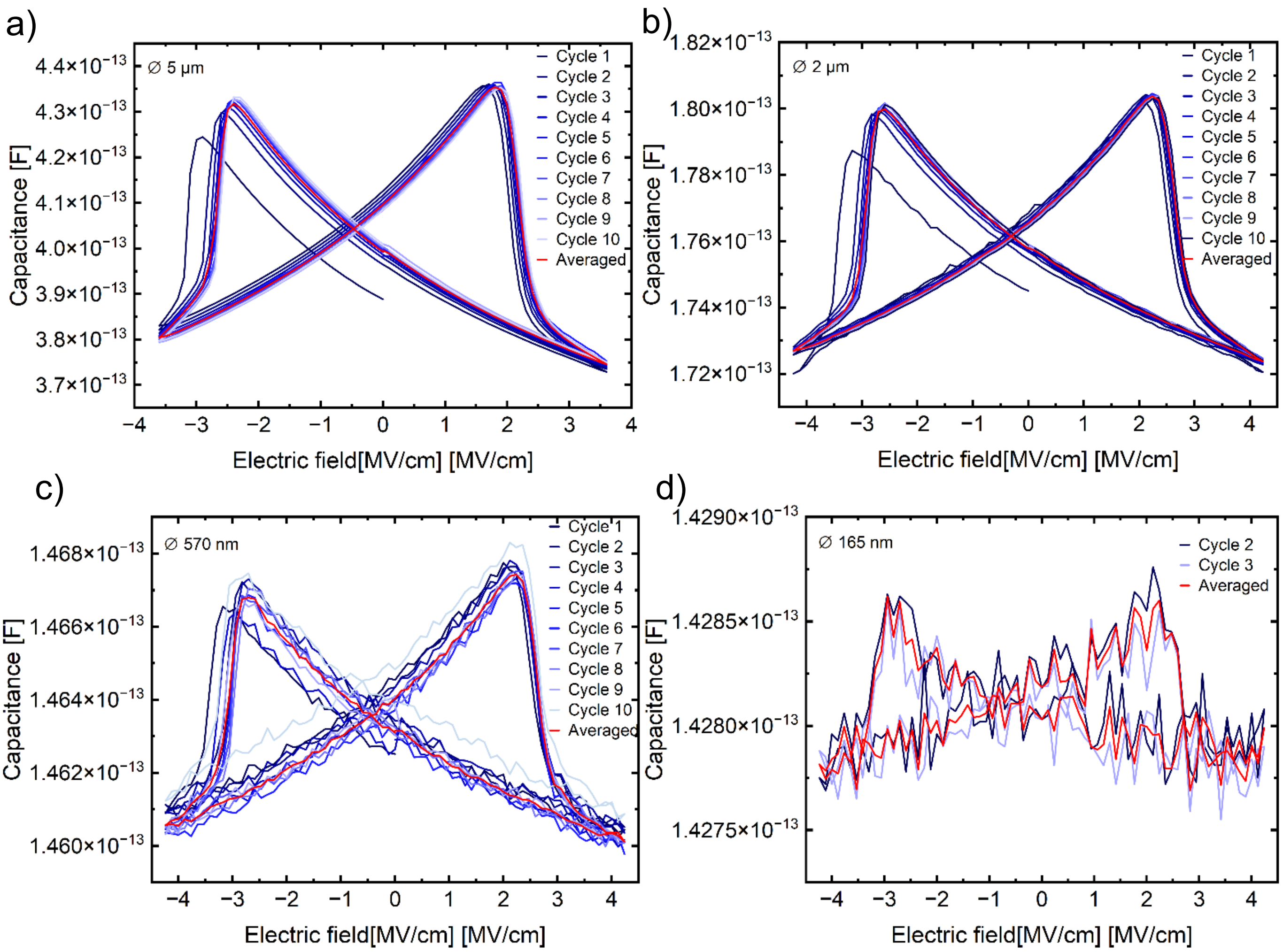

**Figure S4:** Signal-to-noise ratio enhancement via sequential C-V averaging. Raw sequential capacitance-voltage sweeps and their resulting cumulative average for the a) 5 µm, b) 2 µm, c) 570 nm, and d) 165 nm diameter devices. The initial transient wake-up sweep (Cycle 1) is systematically excluded from the evaluation. As the device area shrinks into the sub-micrometer limit, the raw individual cycles become heavily dominated by random instrumental noise, strictly necessitating this sequential averaging protocol to mathematically suppress the statistical variance ($\Delta C_{sys}$) prior to frequency-domain filtering.

To mitigate the remaining high-frequency instrumental noise approaching the attofarad resolution limit, a Fast Fourier Transform (FFT) low-pass filter was applied to the averaged capacitance signal, followed by a multi-peak fitting of asymmetric double sigmoidal functions to accurately resolve the local maxima corresponding to the coercive fields. Figure S5 demonstrates this mathematical recovery of the true ferroelectric "butterfly" hysteresis from the attofarad noise floor. To ensure metrological transparency, **Table S2** catalogues the exact numerical processing parameters, including the number of averaged sweeps, the specific FFT low-pass cutoff frequencies ($f_c$), and the final multi-peak fit qualities ($R^2$), utilized for every device diameter.

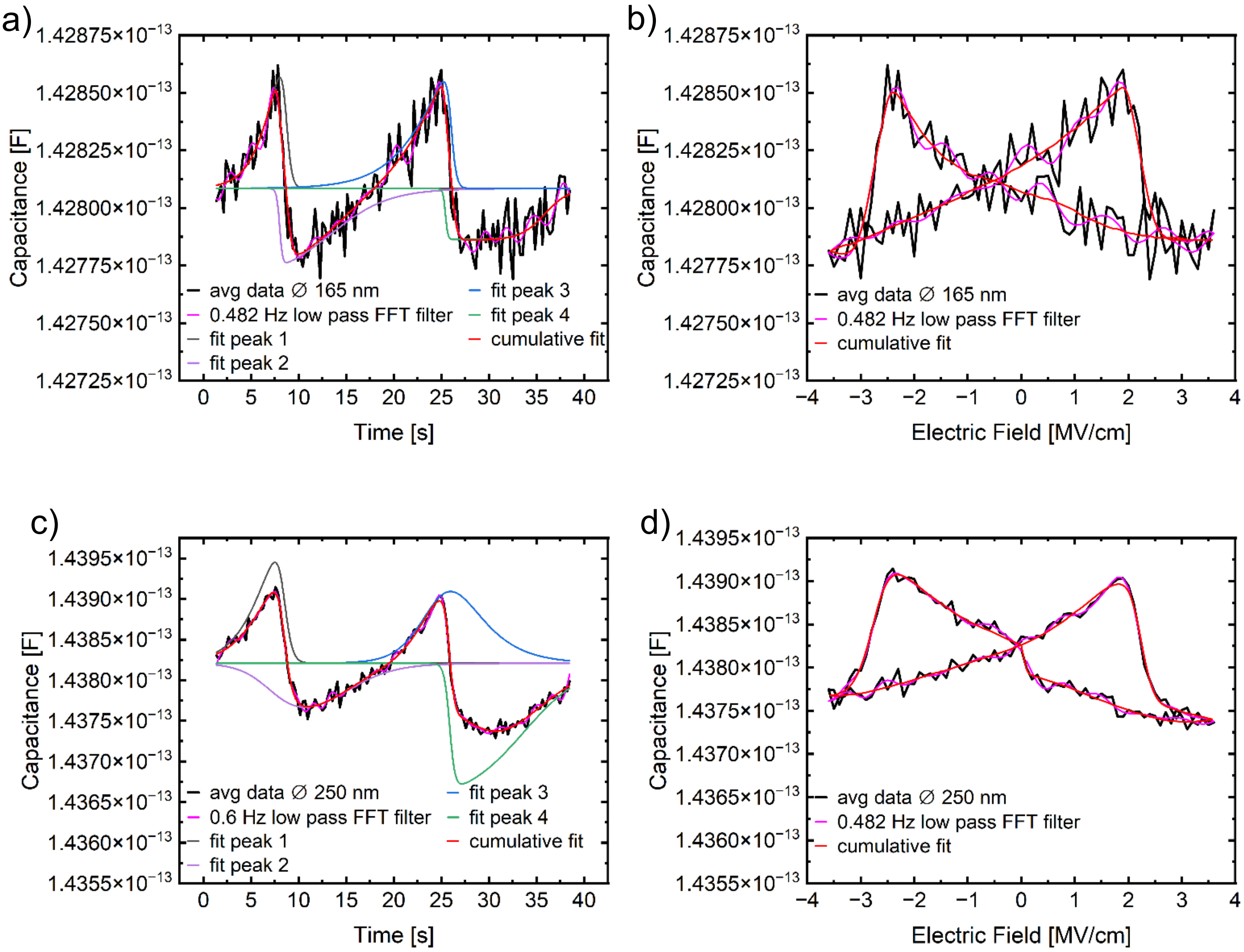

**Figure S5:** Time-to-voltage domain transformation and small-signal parameter extraction. a, c) Raw averaged (black) and FFT low-pass filtered (red) capacitance versus time data for the 165 nm and 250 nm devices, respectively, overlaid with the multi-peak fitting of asymmetric double sigmoidal functions utilized to isolate the intrinsic switching transients. b, d) The identically filtered datasets plotted against the applied electric field, demonstrating the mathematical recovery of the true ferroelectric "butterfly" hysteresis from the attofarad noise floor without introducing phase distortion to the displacement current.

**Table S2:** Comprehensive data curation and statistical variance parameters. Explicit numerical tracking of the sequential averaging, Fast Fourier Transform (FFT) low-pass filtering thresholds ($f_c$), and multi-peak fitting qualities ($R^2$) applied across the dimensional crossover. The relative statistical error bounds the standard error of the mean ($\mathrm{SEM}$) normalized against the intrinsic device capacitance across the full voltage sweep. This mathematically verifies that the isolated noise floor remains securely below the systematic instrumental accuracy limit.

| Device Diameter | Number of Averages | Min Relative Statistical Error [%] | Max Relative Statistical Error [%] | FFT Filter applied? | FFT Low Pass Cutoff ($f_c$) [Hz] | Peak Fit applied? | Cumulative Fit Quality ($R^2$) |
|---|---|---|---|---|---|---|---|
| 51.7 µm | 9 | 0.05 | 1.65 | no | | no | |
| 41.9 µm | 9 | 0.05 | 1.65 | no | | no | |
| 31.9 µm | 9 | 0.05 | 1.65 | no | | no | |
| 21.9 µm | 9 | 0.05 | 1.67 | no | | no | |
| 11.9 µm | 9 | 0.04 | 1.56 | no | | no | |
| 5.7µm | 9 | 0.07 | 1.61 | no | | no | |
| 2 µm | 9 | 0.04 | 1.43 | no | | no | |
| 1.03 µm | 6 | 0.42 | 2.53 | no | | no | |
| 570 nm | 8 | 0.40 | 2.57 | yes | 0.9 | no | |
| 400 nm | 6 | 0.26 | 1.81 | yes | 0.7 | yes | 0.998 |
| 380 nm | 8 | 0.34 | 1.72 | yes | 0.8 | yes | 0.995 |
| 330 nm | 6 | 0.38 | 1.48 | yes | 0.482 | yes | 0.997 |
| 250 nm | 4 | 2.26 | 4.54 | yes | 0.6 | yes | 0.994 |
| 165 nm | 2 | 0.12 | 7.10 | yes | 0.482 | yes | 0.978 |

## Supplementary Note 9: $R_p$ Data processing

The parallel equivalent resistance ($R_\mathrm{p}$) was extracted concurrently with the capacitance. Because $R_\mathrm{p}$ and $C_\mathrm{p}$are mathematically coupled through the LCR meter's complex impedance measurement ($Z = R_\mathrm{p}||\frac{1}{j\omega c}$), asymmetric signal processing would introduce severe phase distortion. Therefore, the exact same sequential averaging protocol and precise FFT low-pass cutoff frequencies ($f_\mathrm{c}$) detailed in **Table S2** were applied simultaneously to the raw $R_\mathrm{p}$ time-domain waveforms. This symmetric filtering ensures strict phase synchronicity between the real and imaginary impedance components, mathematically validating the subsequent conductance scaling and dynamic dielectric loss extractions.

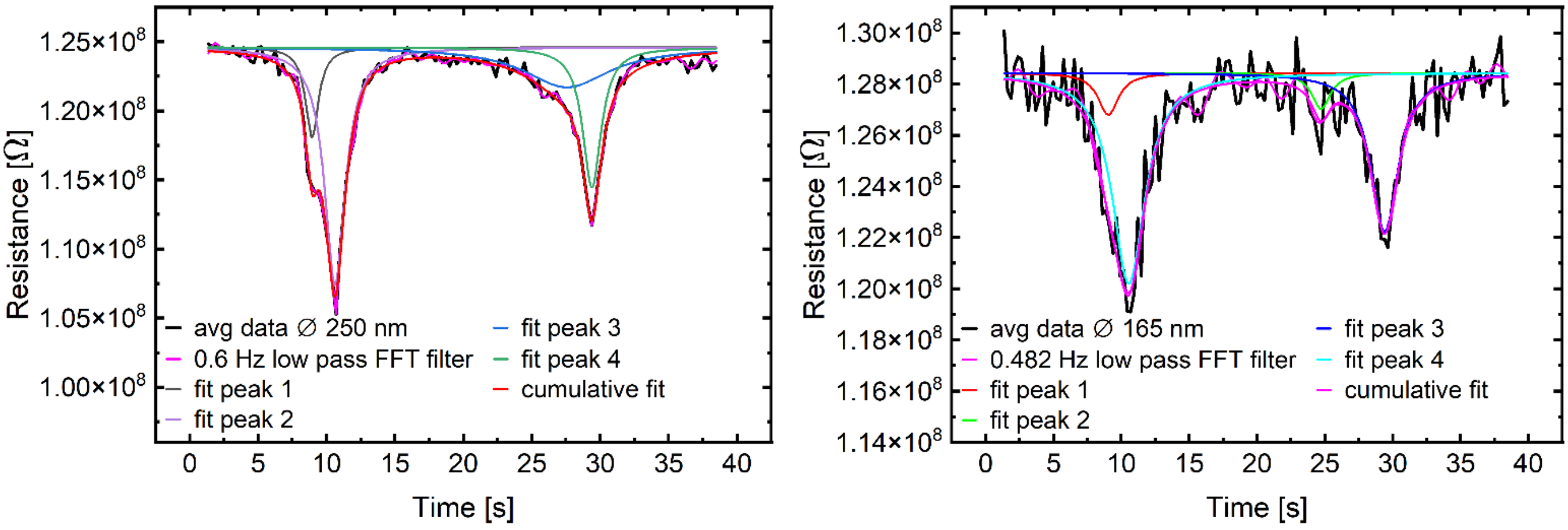

**Figure S6:** Phase-synchronous impedance filtering at the scaling limit. Time-domain transients of the parallel equivalent resistance ($R_p$) overlaid with the multi-peak sigmoidal fits for the (a) 250 nm and (b) 165 nm diameter devices. To prevent phase distortion during de-embedding, the exact same Fast Fourier Transform (FFT) low-pass cutoff frequencies utilized for the capacitance data (0.6 Hz and 0.482 Hz, respectively) were applied to these averaged resistive waveforms. This symmetric protocol validates the mathematical fidelity of the intrinsic dynamic loss extraction.

## Supplementary Note 10: J-E Data post processing

Large-signal dynamic J-E loops were acquired utilizing a standard Positive Up Negative Down (PUND) protocol to strictly isolate the intrinsic displacement current from the divergent background resistive leakage. The sequence applies a 50 µs unipolar triangular pulse for the active switching phase, immediately followed by identical non-switching pulses. By mathematically subtracting the non-switching current response, which contains only the linear dielectric charging and parallel resistive perimeter leakage, from the initial switching response, the true ferroelectric polarization reversal is isolated. In the deep sub-micrometer limit, this leakage-corrected transient was subsequently subjected to the FFT filtering and sequential exponential fitting framework detailed in **Figure S7** of the main text to extract the unmasked dynamic coercive fields.

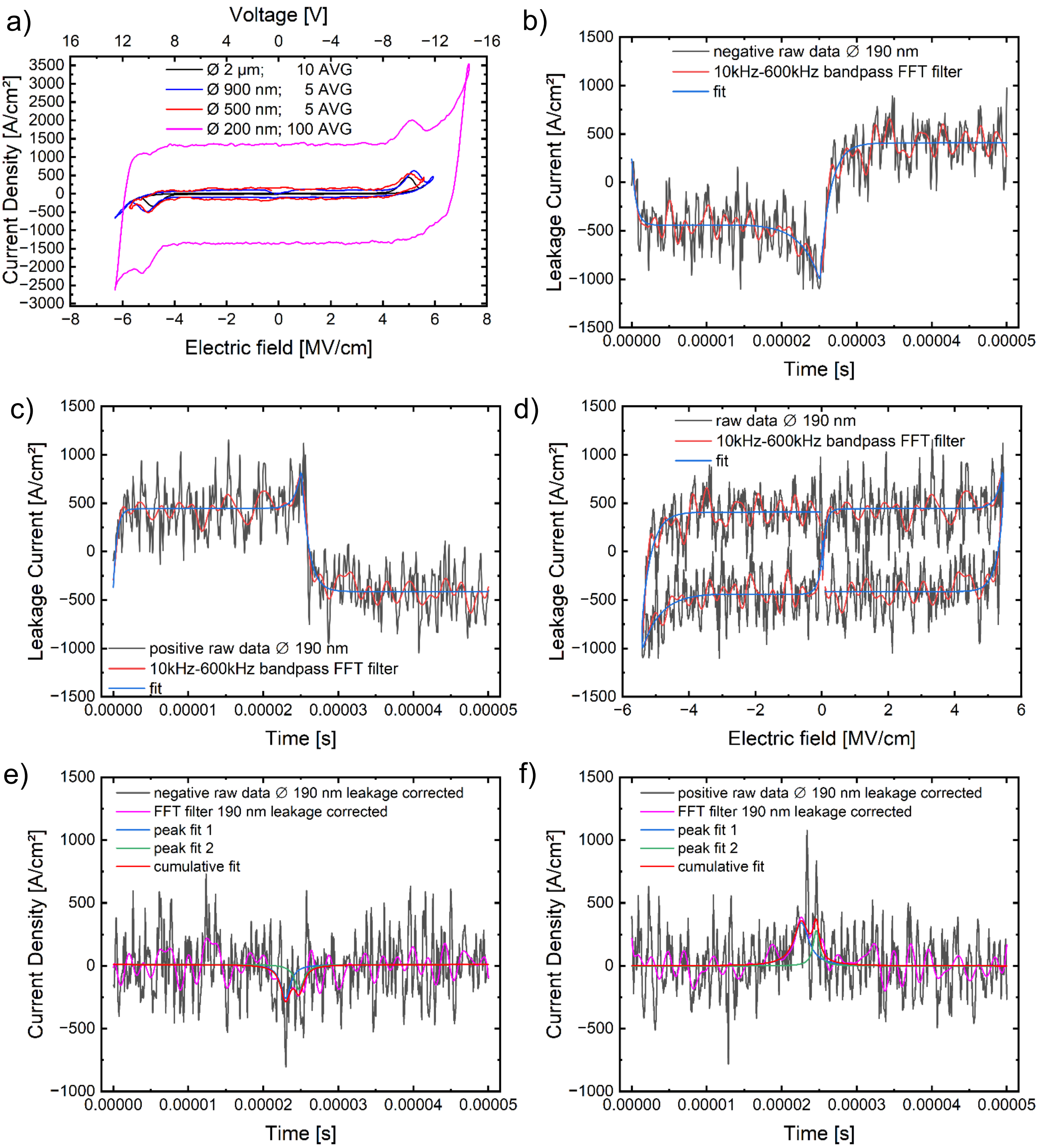

**Figure S7:** Dynamic leakage compensation and J-E hysteresis recovery. a) Large-signal dynamic J-E loops at 30 kHz acquired across the dimensional scaling range. The divergence of the uncompensated current at higher fields necessitates rigorous leakage extraction. b-d) Time-domain and electric-field-domain profiles of the raw leakage current for the 190 nm device. To isolate the non-switching perimeter leakage from the ferroelectric displacement current, a 10 kHz-600kHz bandpass FFT filter was applied to the raw transient, followed by an analytical fit (blue line) to establish the continuous background conduction profile. e, f) The final leakage-corrected current density transients for the negative and positive polarities, respectively. By mathematically subtracting the fitted background leakage, the true intrinsic ferroelectric switching peaks (green/purple) are isolated and recovered from the noise-dominated raw signal, enabling the accurate extraction of the coercive fields in the deep sub-micrometer limit.

## 4.2 SI References

1 Shi, C., Luu, D.K., Yang, Q., Liu, J., Chen, J., Ru, C., Xie, S., Luo, J., Ge, J., Sun, Y. (2016) Recent advances in nanorobotic manipulation inside scanning electron microscopes. *Microsyst Nanoeng*, **2** (1), 16024.

2 Wang, R., Zhou, J., Zheng, S., Zhu, F., Sun, W., Xu, H., Li, B., Liu, Y., Hao, Y., Han, G. (2025) Cryogenic Ferroelectric Behavior of Wurtzite Ferroelectrics.

3 Jabbari, I., Baira, M., Maaref, H., Mghaieth, R. (2020) Evidence of Poole-frenkel and Fowler-Nordheim tunneling transport mechanisms in leakage current of (Pd/Au)/Al0.22Ga0.78N/GaN heterostructures. *Solid State Communications*, **314-315**, 113920.

4 Test Methods for Determining Average Grain Size, E04 Committee, West Conshohocken, PA.

5 Kirchhoff, G.R. (1877) Zur Theorie des Condensators, 144–162.

6 Palmer, H.B. (1937) The Capacitance of a Parallel-Plate Capacitor by the Schwartz-Christoffel Transformation. *Trans. Am. Inst. Electr. Eng.*, **56** (3), 363–366.

7 Nishiyama, H. and Nakamura, M. (1994) Form and capacitance of parallel-plate capacitors. *IEEE Trans. Comp., Packag., Manufact. Technol. A*, **17** (3), 477–484.

8 S. Hudlet, M. Saint Jean, C. Guthmann, J. Berger (1998) Evaluation of the capacitive force between an atomic force microscopy tip and a metallic surface. *The European Physical Journal B - Condensed Matter and Complex Systems*.

9 Jackson, J.D. (2009) *Classical electrodynamics*, 3rd edn, Wiley, Hoboken, NY.

10 Dutta, S., Sankaran, K., Moors, K., Pourtois, G., van Elshocht, S., Bömmels, J., Vandervorst, W., Tőkei, Z., Adelmann, C. (2017) Thickness dependence of the resistivity of platinum-group metal thin films. *Journal of Applied Physics*, **122** (2).

11 Hewlett-Packard Japan, LTD. (1998) HP 4284A Precision LCR Meter - Operation manual.